\documentclass[twocolumn]{aastex7}
\usepackage{booktabs}
\usepackage{multirow}
\usepackage{graphicx}
\usepackage{amsmath} 

\providecommand{\teff}{\ensuremath{T_{\rm eff}}} 
\providecommand{\msun}{\ensuremath{\,{\rm M_{\odot}}}} 
\providecommand{\rsun}{\ensuremath{\,{\rm R_{\odot}}}} 
\newcommand{\vsini}{\ensuremath{v\sin i_\star}} 

\shorttitle{Obliquity Measurements of TOI-560~b and TOI-5082~b}
\shortauthors{Zhang et al.}

\begin{document}

\title{Spin-Orbit Alignment of Two Neptune-size Planets Younger than 500 Myr: TOI-560~b and TOI-5082~b}

\author[0009-0004-0455-2424]{Elina Y. Zhang}
\email{yuchenzh@hawaii.edu}
\affiliation{Institute for Astronomy, University of Hawaii, 2680 Woodlawn Drive, Honolulu, HI 96822 USA}

\author[0000-0002-8958-0683]{Fei Dai}
\email{fdai@hawaii.edu}
\affiliation{Institute for Astronomy, University of Hawaii, 2680 Woodlawn Drive, Honolulu, HI 96822 USA}

\author[0000-0001-8638-0320]{Andrew W. Howard}
\email{ahoward@caltech.edu}
\affiliation{Department of Astronomy, California Institute of Technology, Pasadena, CA 91125, USA}

\author[0000-0003-1312-9391]{Samuel P. Halverson}
\email{samuel.halverson@jpl.nasa.gov}
\affiliation{Jet Propulsion Laboratory, California Institute of Technology, 4800 Oak Grove Drive, Pasadena, CA 91109, USA}

\author[0000-0002-0531-1073]{Howard Isaacson}
\email{hisaacson@berkeley.edu}
\affiliation{{Department of Astronomy,  University of California Berkeley, Berkeley CA 94720, USA}}

\author[0000-0003-3856-3143]{Ryan A. Rubenzahl}
\email{rrubenzahl@flatironinstitute.org}
\affiliation{Center for Computational Astrophysics, Flatiron Institute, 162 Fifth Avenue, New York, NY 10010, USA}

\author[0000-0003-3860-6297]{Huan-Yu Teng}
\email{huanyu.teng.astro@gmail.com}
\affiliation{Korea Astronomy and Space Science Institute, 776 Daedeok-daero, Yuseong-gu, Daejeon 34055, Republic of Korea}
\affiliation{National Astronomical Observatories, Chinese Academy of Sciences, Beijing 100101, China}

\author[0000-0002-0376-6365]{Xian-Yu Wang}
\email{xwa5@iu.edu}
\affiliation{Department of Astronomy, Indiana University, Bloomington, IN 47405, USA}

\author[0000-0002-7846-6981]{Songhu Wang}
\email{sw121@iu.edu}
\affiliation{Department of Astronomy, Indiana University, Bloomington, IN 47405, USA}

\author[0000-0003-3244-5357]{Daniel Hey}
\email{dhey@hawaii.edu}
\affiliation{Institute for Astronomy, University of Hawaii, 2680 Woodlawn Drive, Honolulu, HI 96822 USA}

\author[0000-0001-8832-4488]{Daniel Huber}
\email{huberd@hawaii.edu}
\affiliation{Institute for Astronomy, University of Hawaii, 2680 Woodlawn Drive, Honolulu, HI 96822 USA}
\affiliation{Sydney Institute for Astronomy (SIfA), School of Physics, University of Sydney, NSW 2006, Australia}
\author[0000-0003-3504-5316]{Benjamin J. Fulton}
\email{bjfulton@ipac.caltech.edu}
\affiliation{NASA Exoplanet Science Institute/Caltech-IPAC, MC 314-6, 1200 E California Blvd, Pasadena, CA 91125, USA}

\author[0000-0001-8342-7736]{Jack Lubin}
\email{jblubin@astro.ucla.edu}
\affiliation{Department of Physics \& Astronomy, University of California Los Angeles, Los Angeles, CA 90095, USA}

\author[0000-0002-9305-5101]{Luke B. Handley}
\email{lhandley@caltech.edu}
\affiliation{Department of Astronomy, California Institute of Technology, Pasadena, CA 91125, USA}

\author[0000-0002-8965-3969]{Steven Giacalone}
\email{giacalone@astro.caltech.edu}
\affiliation{Department of Astronomy, California Institute of Technology, Pasadena, CA 91125, USA}

\author[0000-0003-0967-2893]{Erik A. Petigura}
\email{petigura@astro.ucla.edu}
\affiliation{Department of Physics \& Astronomy, University of California Los Angeles, Los Angeles, CA 90095, USA}

\author[0000-0002-3725-3058]{Lauren M. Weiss}
\email{lweiss4@nd.edu}
\affiliation{Department of Physics and Astronomy, University of Notre Dame, Notre Dame, IN, 46556, USA}

\author[0000-0002-5812-3236]{Aaron Householder}
\email{aaron593@mit.edu}
\affiliation{Department of Earth, Atmospheric and Planetary Sciences, Massachusetts Institute of Technology, Cambridge, MA 02139, USA}
\affiliation{Kavli Institute for Astrophysics and Space Research, Massachusetts Institute of Technology, Cambridge, MA 02139, USA}

\author[0000-0002-6406-1924]{Casey Y. Lam}
\email{clam@carnegiescience.edu}
\affiliation{Observatories of the Carnegie Institution for Science, 813 Santa Barbara Street, Pasadena, CA 91101, USA}


\author[0000-0002-4290-6826]{Judah Van Zandt}
\email{judahvz@astro.ucla.edu}
\affiliation{Department of Physics \& Astronomy, University of California Santa Barbara, Santa Barbara, CA 93106, USA}


\author[0009-0004-4454-6053]{Steve R. Gibson}
\email{sgibson@caltech.edu}
\affil{Caltech Optical Observatories, Pasadena, CA, 91125, USA}

\author{Kodi Rider}
\email{kodi.rider@ssl.berkeley.edu}
\affil{Space Sciences Laboratory, University of California Berkeley, Berkeley, CA 94720, USA}

\author[0000-0001-8127-5775]{Arpita Roy}
\email{arpita308@gmail.com}
\affiliation{Astrophysics \& Space Institute, Schmidt Sciences, New York, NY 10011, USA} 

\author[0000-0002-6525-7013]{Ashley Baker}
\email{abaker@caltech.edu}
\affil{Caltech Optical Observatories, Pasadena, CA, 91125, USA}

\author[0009-0002-2419-8819]{Jerry Edelstein}
\email{jerrye@ssl.berkeley.edu}
\affil{Space Sciences Laboratory, University of California Berkeley, Berkeley, CA 94720, USA}

\author{Christopher L. Smith} 
\email{christopher.smith@berkeley.edu}
\affil{Space Sciences Laboratory, University of California Berkeley, Berkeley, CA 94720, USA}

\author[0000-0002-6092-8295]{Josh Walawender}
\email{jwalawender@keck.hawaii.edu}
\affiliation{W. M. Keck Observatory, 65-1120 Mamalahoa Hwy, Waimea, HI 96743}

\author[0000-0002-4265-047X]{Joshua N. Winn}
\email{jnwinn@princeton.edu}
\affiliation{Department of Astrophysical Sciences, Princeton University, 4 Ivy Lane, Princeton, NJ 08544, USA}

\begin{abstract}
\noindent
Stellar obliquity measurements provide a direct probe of planetary system dynamics, but remain sparse for Neptune-size planets, particularly at young ages. We present Rossiter--McLaughlin measurements for two young Neptune-size planets, TOI-560\,b and TOI-5082\,b, using time-resolved Keck Planet Finder (KPF) spectroscopy and joint modeling with \textit{TESS} transit photometry. We measure sky-projected obliquities of $\lambda_b = -25 \pm 16^\circ$ for TOI-560\,b and $\lambda_b = 19^{+17}_{-13}{}^\circ$ for TOI-5082\,b. Combining these constraints with stellar rotation periods and spectroscopic estimates of $\vsini$, we obtain 95\% upper limits of $\psi < 69.7^\circ$ and $\psi < 50.5^\circ$, respectively. Both systems are therefore consistent with low-to-moderate true obliquities and show no evidence of strong spin–orbit misalignment. With ages of $480 \pm 190$ Myr for TOI-560\,b and $180 \pm 9$ Myr for TOI-5082\,b (a likely member of the CRIUS197 stellar association), the systems are in a key evolutionary phase when post-disk dynamical processes such as secular interactions may begin to manifest. Nevertheless, both systems remain consistent with low obliquity. In the broader context of young systems with existing measurements, these results support an emerging picture in which Neptune-size planets at $\lesssim 1$\,Gyr are commonly found in low-obliquity configurations.
\end{abstract}

\keywords{Transits (1711), Exoplanets (498), Exoplanet dynamics (490), Planet hosting stars (1242), Planetary system formation (1257), Young stellar objects (1834)}

\section{Introduction} 
\label{sec:intro}
Stellar obliquity measurements provide a direct probe of the dynamical evolution of planetary systems. The stellar obliquity, $\psi$, is defined as the angle between the orbit normal of the planet ($\hat{n}_{\rm orb}$) and the spin direction of the host star ($\hat{n}_{\star}$). Observationally, one often measures the sky-projected obliquity, $\lambda$, defined as the angle between their projections onto the plane of the sky. The Rossiter--McLaughlin or RM effect enables measurements of $\lambda$ through anomalous Doppler shifts observed during planetary transits \citep[RM;][]{Rossiter1924,McLaughlin1924,Queloz2000,GaudiWinn2007}. With additional constraints on the stellar inclination (e.g., from $\vsini$ and stellar rotation), the true obliquity $\psi$ can be inferred \citep[e.g.,][]{Winn2005,MasudaWinn2020}.

Planets with sizes between Earth and Neptune seem to be very common outcomes of planet formation \citep{Fressin, Petigura}, yet they have no Solar System analog. Their properties encode the coupled effects of formation, migration, and evolution, making them a key population for testing planet formation theories \citep{Winn,Zhu,Owen}. In addition to their atmospheric evolution, stellar obliquity measurements for these planets provide a complementary probe of their dynamical histories \citep{Albrecht2022}. While hot Jupiters have historically dominated RM studies due to their large signals, extending obliquity measurements to Neptune-size planets is essential for determining whether spin--orbit alignment is a generic outcome of planet formation or whether dynamical excitation is common across a broader range of planet sizes. Ultra-stable, high-resolution spectrographs such as the Keck Planet Finder \citep{Gibson2024,Gibson2020} facilitate RM measurements of smaller planets around bright host stars. Recent measurements include independent studies of the young Neptune K2-25~b \citep{Stefansson2020,Gaidos2020}, as well as measurements for systems spanning warm Neptunes and compact sub-Saturn architectures \citep[e.g.,][]{GJ436b,GJ3470b,Zak2024,Radzom2024,Mantovan2024}. These efforts are being expanded through systematic surveys of Neptune-size planets, such as the ATREIDES \citep{Bourrier2025ATREIDES}, KPF-SLOPE \citep{Handley}, and the POSEIDON programs \citep{Espinoza-Retamal2026}.

Young planetary systems are especially informative because both planetary evolution and dynamical evolution may be expected to be most rapid early in a system’s lifetime \citep{Izidoro2017,Schmidt,Dai2024}. Close-in Neptune-size planets may undergo substantial radius evolution over $\sim$10–1000~Myr as their envelopes cool and contract and as escape processes remove volatile gases, potentially driving planets across the radius valley \citep[e.g.,][]{OwenWu2017,Ginzburg2018,Gupta2019,Owen2019}. At the same time, the dynamical architectures of planetary systems may evolve over a wide range of timescales \citep{DawsonJohnson2018,Triaud2018}. Mechanisms that excite spin–orbit misalignment span orders of magnitude in time, from primordial disk tilting during the gas-disk phase ($\lesssim 10$~Myr; \citealt{Haisch2001,Mamajek2009,Batygin2012,Lai2014}) to Kozai–Lidov cycles ($10^{4}$–$10^{8}$~yr; \citealt{Fabrycky2007,Wu2007,Naoz2016}) and secular interactions among planets ($10^{7}$–$10^{9}$~yr; \citealt{WuLithwick2011,Petrovich2015}). Planet–planet scattering can also excite eccentricities and inclinations over Myr–Gyr timescales after disk dispersal \citep{Rasio1996,Chatterjee2008,Juric2008}. Conversely, tidal dissipation may damp obliquities and eccentricities over Gyr timescales \citep{Winn2010,Albrecht2012}. Measuring stellar obliquities for systems with well-constrained ages therefore provides a direct means of distinguishing between these pathways. In particular, obliquity measurements of young and intermediate-age systems can test whether spin–orbit misalignment is established early or emerges gradually over hundreds of Myr. Although RM measurements for young and/or active hosts remain observationally challenging, a growing sample now probes this regime, including TIDYE-1~b \citep{Barber2025}, AU~Mic~b \citep{Palle2020}, DS~Tuc~Ab \citep{Zhou2020,Benatti2021}, V1298~Tau~b \citep{Johnson2022}, TIC~150070085~b \citep{Barber2026}, K2-25~b \citep{Stefansson2020,Gaidos2020}, TOI-5398~b \citep{Radzom2024,Mantovan2024}, and TOI-837~b \citep{Mantovan2026}. Building samples that span a range of system ages is crucial for establishing a time-resolved picture of dynamical evolution. Neptune-size planets in young systems therefore provide a uniquely powerful laboratory, linking atmospheric evolution with dynamical histories through joint constraints on planetary structure and spin–orbit architecture.

In this paper, we present stellar obliquity measurements for two Neptune-size planets, TOI-560~b and TOI-5082~b, based on Rossiter--McLaughlin observations obtained with the Keck Planet Finder (KPF) and joint modeling with \textit{TESS} transit photometry. These systems occupy a particularly informative age range, with TOI-5082~b ($\sim$200~Myr, if associated with CRIUS197; \citealt{Robbins2023,Gagne2026}) and TOI-560~b ($\sim$500~Myr; \citealt{Barragan2022_TOI560}) probing intermediate stages of planetary evolution when dynamical processes may still be active. Together, these systems provide an opportunity to investigate how spin--orbit architectures evolve with time for Neptune-size planets. Section~\ref{sec:observation} describes the \textit{TESS} light curves and the KPF in-transit spectroscopic sequences. Section~\ref{sec:stellar_para} presents our determination of the host-star properties using spectroscopic constraints and isochrone modeling. In Section~\ref{sec:analysis}, we perform a joint analysis of the \textit{TESS} photometry and RM radial velocities to infer the sky-projected obliquities, and we incorporate stellar rotation constraints to inform the three-dimensional spin--orbit geometry. We discuss the implications of these measurements for the spin--orbit architectures of Neptune-sized planets and early system evolution in Section~\ref{sec:discussion}, and summarize our conclusions in Section~\ref{sec:summary}.

\section{Observations} \label{sec:observation}

\subsection{{\it TESS} Light Curves}
\label{sec:tess_lc}

TOI-560 was observed by the {\it Transiting Exoplanet Survey Satellite} ({\it TESS}) in Sectors 8, 31, 64, and 88, while TOI-5082 was observed in Sectors 44, 45, 71, and 72. For both systems, we analyzed 2-minute cadence light curves produced by the {\it TESS} Science Processing Operations Center \citep[SPOC;][]{Jenkins2016}. All {\it TESS} data used in this work are publicly available from the Mikulski Archive for Space Telescopes (MAST)\footnote{\url{https://mast.stsci.edu/portal/Mashup/Clients/Mast/Portal.html}} and can be accessed via the MAST data set DOI: \dataset[10.17909/yf0b-wj43]{https://doi.org/10.17909/yf0b-wj43}.

We used the Presearch Data Conditioning Simple Aperture Photometry (PDCSAP) fluxes \citep{Stumpe2014, Smith2012}, which are corrected for common instrumental systematics while preserving astrophysical signals such as planetary transits. The light curves were processed using the \texttt{lightkurve} Python package \citep{lightkurve2018}. As an initial cleaning step, we removed outlier points exceeding a 10$\sigma$ threshold to eliminate spurious measurements.

To prevent transit signals from being distorted during detrending, we applied a transit mask based on the known orbital ephemerides of each planet. Long-term trends and low-frequency variability arising from stellar activity, spacecraft systematics, and background contamination were then removed using a Savitzky–Golay filter \citep{Savitzky1964} implemented within \texttt{lightkurve}, with the in-transit data excluded from the filtering process. This procedure preserves the integrity of the transit profiles while maintaining accurate photometric uncertainties for subsequent modeling.

\subsection{KPF Rossiter--McLaughlin Observations}

Measurements of stellar obliquities have historically focused on hot Jupiter systems, where the Rossiter--McLaughlin (RM) signal is largest due to its dependence on the squared planet-to-star radius ratio, $(R_{\rm p}/R_\star)^2$. The combination of high spectral resolution and high instrumental stability offered by KPF \citep{Gibson2016, Gibson2018, Gibson2020, Gibson2024} facilitates RM studies of smaller planets. KPF is a fiber-fed, high-resolution echelle spectrograph installed at the W.\ M.\ Keck Observatory, covering wavelengths from 445 to 870~nm across two spectral channels with a median resolving power of $R \approx 98{,}000$. The instrument is optimized for precision Doppler spectroscopy and routinely achieves single-measurement radial velocity precision at or below the 1~m\,s$^{-1}$ level. Wavelength calibration is provided by ThAr and UNe emission lamps, supplemented by a laser frequency comb from Menlo Systems \citep{Rubenzahl2023}.

We observed the Rossiter--McLaughlin effect for TOI-560 and TOI-5082 using KPF during planetary transit events on UTC 2024 December 16 and UTC 2024 November 26, respectively. The TOI-560 observations were obtained during a transit of planet~b and consisted of 47 consecutive exposures, achieving a typical signal-to-noise ratio of approximately 150 per pixel at 550~nm. Similarly, the RM sequence for TOI-5082 comprised 46 exposures obtained during a transit of planet~b, with a typical signal-to-noise ratio of approximately 340 per pixel at 550~nm. Each exposure had an integration time of 300 seconds.

The spectra were processed using the KPF Data Reduction Pipeline \citep[DRP;][]{Gibson2024, Gibson2020}, which performs bias subtraction, flat-fielding, spectral extraction, and wavelength calibration. Radial velocities were derived using the cross-correlation function (CCF) technique with numerical masks based on those developed for {\tt ESPRESSO} \citep{Pepe2013}. The resulting RV time series used in the RM analysis are listed in Table~\ref{tab:KPF_data_parallel}.

\begin{deluxetable}{@{}ccc|ccc@{}}
\tablecaption{KPF Rossiter--McLaughlin Data for TOI-560 and TOI-5082. 
Time is given in BJD$-$2457000. \label{tab:KPF_data_parallel}}
\tablehead{
\multicolumn{3}{c}{TOI-560} & \multicolumn{3}{c}{TOI-5082} \\
\colhead{\shortstack{Time \\ ($\Delta$BJD)}} & 
\colhead{\shortstack{RV \\ (km/s)}} & 
\colhead{\shortstack{$\sigma_{\rm RV}$ \\ (m/s)}} &
\colhead{\shortstack{Time \\ ($\Delta$BJD)}} & 
\colhead{\shortstack{RV \\ (km/s)}} & 
\colhead{\shortstack{$\sigma_{\rm RV}$ \\ (m/s)}}
}
\startdata
3660.9475 & 20.83042 & 0.87 & 3640.9231 & 42.90057 & 0.91 \\
3660.9515 & 20.82671 & 0.91 & 3640.9271 & 42.90048 & 0.90 \\
3660.9556 & 20.82931 & 0.80 & 3640.9311 & 42.90051 & 0.87 \\
3660.9596 & 20.82943 & 0.72 & 3640.9352 & 42.90098 & 0.87 \\
3660.9637 & 20.82573 & 0.72 & 3640.9391 & 42.90182 & 0.88 \\
3660.9678 & 20.82996 & 0.68 & 3640.9472 & 42.90331 & 0.85 \\
3660.9718 & 20.82856 & 0.69 & 3640.9515 & 42.90208 & 0.92 \\
3660.9756 & 20.82707 & 0.69 & 3640.9553 & 42.90226 & 0.81 \\
3660.9798 & 20.82769 & 0.71 & 3640.9594 & 42.90142 & 0.83 \\
3660.9838 & 20.82921 & 0.70 & 3640.9632 & 42.90161 & 0.88 \\
3660.9877 & 20.82848 & 0.78 & 3640.9677 & 42.90040 & 0.94 \\
3660.9920 & 20.82642 & 0.89 & 3640.9715 & 42.89979 & 0.86 \\
3660.9959 & 20.83040 & 0.76 & 3640.9757 & 42.90157 & 0.88 \\
3661.0001 & 20.82818 & 0.76 & 3640.9795 & 42.90010 & 0.83 \\
3661.0041 & 20.83039 & 0.72 & 3640.9836 & 42.89950 & 0.82 \\
3661.0081 & 20.82937 & 0.70 & 3640.9875 & 42.89962 & 0.85 \\
3661.0121 & 20.83160 & 0.66 & 3640.9917 & 42.89763 & 0.79 \\
3661.0162 & 20.82910 & 0.65 & 3640.9957 & 42.89857 & 0.81 \\
3661.0201 & 20.83191 & 0.63 & 3640.9997 & 42.89803 & 0.82 \\
3661.0242 & 20.82757 & 0.65 & 3641.0038 & 42.89633 & 0.79 \\
3661.0282 & 20.82857 & 0.64 & 3641.0078 & 42.89650 & 0.80 \\
3661.0322 & 20.82838 & 0.63 & 3641.0118 & 42.89708 & 0.81 \\
3661.0362 & 20.82873 & 0.65 & 3641.0160 & 42.89594 & 0.79 \\
3661.0411 & 20.82517 & 0.69 & 3641.0198 & 42.89491 & 0.79 \\
3661.0450 & 20.82670 & 0.68 & 3641.0239 & 42.89469 & 0.81 \\
3661.0492 & 20.82534 & 0.61 & 3641.0280 & 42.89469 & 0.77 \\
3661.0531 & 20.82443 & 0.63 & 3641.0320 & 42.89526 & 0.82 \\
3661.0570 & 20.82736 & 0.64 & 3641.0360 & 42.89418 & 0.92 \\
3661.0611 & 20.82634 & 0.66 & 3641.0402 & 42.89816 & 0.94 \\
3661.0653 & 20.82423 & 0.65 & 3641.0441 & 42.89761 & 0.78 \\
3661.0692 & 20.82488 & 0.65 & 3641.0480 & 42.89800 & 0.83 \\
3661.0732 & 20.82642 & 0.73 & 3641.0523 & 42.89829 & 0.91 \\
3661.0772 & 20.82481 & 0.68 & 3641.0562 & 42.89701 & 0.83 \\
3661.0812 & 20.82577 & 0.61 & 3641.0603 & 42.89663 & 0.77 \\
3661.0853 & 20.82517 & 0.65 & 3641.0643 & 42.89697 & 0.75 \\
3661.0893 & 20.82509 & 0.65 & 3641.0685 & 42.89663 & 0.79 \\
3661.0934 & 20.82569 & 0.73 & 3641.0724 & 42.89644 & 0.74 \\
3661.0975 & 20.82157 & 0.90 & 3641.0763 & 42.89736 & 0.76 \\
3661.1015 & 20.82647 & 0.92 & 3641.0804 & 42.89807 & 0.73 \\
3661.1058 & 20.82249 & 0.82 & 3641.0845 & 42.89811 & 0.73 \\
3661.1096 & 20.82414 & 0.66 & 3641.0884 & 42.89664 & 0.74 \\
3661.1137 & 20.82487 & 0.65 & 3641.0926 & 42.89669 & 0.73 \\
3661.1176 & 20.82490 & 0.65 & 3641.0965 & 42.89663 & 0.74 \\
3661.1223 & 20.82494 & 0.69 & 3641.1005 & 42.89667 & 0.78 \\
3661.1263 & 20.82314 & 0.69 & 3641.1045 & 42.89572 & 0.85 \\
3661.1303 & 20.82506 & 0.76 & 3641.1087 & 42.89617 & 0.76 \\
3661.1344 & 20.82353 & 0.83 & -- & -- & -- \\
\enddata
\end{deluxetable}

\section{Stellar Parameters from Isochrone Modeling}
\label{sec:stellar_para}

We derived the fundamental stellar parameters of TOI-560 and TOI-5082 using the \texttt{isoclassify} software \citep{isoclassify_code}, which performs isochrone-based inference by combining spectroscopic, photometric, and astrometric constraints within a Bayesian framework \citep{Huber2017,Berger2020,Berger2023}. In particular, we employed the \texttt{isoclassify} Grid Mode, which derives posterior probability distributions for stellar parameters through direct integration of stellar isochrones, given a set of observational constraints and priors. This approach enables a self-consistent determination of stellar properties by comparing the observed parameters to grids of stellar evolutionary models.

For each target, we adopted spectroscopic measurements of the effective temperature ($T_{\rm eff}$) and metallicity ([Fe/H]) as inputs to the isochrone analysis. Broadband photometry and Gaia parallax measurements were also incorporated to constrain the stellar luminosity and radius \citep{Gaia2016,Gaia2023}. The Gaia parallaxes were corrected for the systematic zero-point offset following \citet{Lindegren2021}. Photometric uncertainties and parallax errors were propagated throughout the analysis.

Isochrone fitting was performed using the MESA Isochrones and Stellar Tracks \citep[MIST;][]{MIST}, as implemented in the \texttt{isoclassify} Grid Mode. For each point in the model grid, the likelihood of the observed parameters was computed assuming Gaussian uncertainties, and posterior probability distributions were obtained by marginalizing over stellar mass, age, metallicity, distance, and extinction. Interstellar extinction was treated as a free parameter and constrained using three-dimensional dust maps accessed via the \texttt{mwdust} package \citep{Bovy2016}.

From the resulting posterior distributions, we adopted the median values and central 68\% credible intervals for the stellar mass, radius, surface gravity, luminosity, stellar density, age, and distance. The final stellar parameters for TOI-560 and TOI-5082 are reported in Table~\ref{tab:stellar_params} and are used throughout this work, including in the modeling of the Rossiter--McLaughlin effect and the inference of the stellar obliquity.

We note that \texttt{isoclassify} does not explicitly account for systematic uncertainties arising from differences between stellar evolutionary model grids. As discussed by \citet{Tayar2022}, such systematics may contribute uncertainties at the level of $\sim$2\% in $T_{\rm eff}$, $\sim$4\% in stellar radius, and $\sim$5\% in stellar mass.

\begin{deluxetable*}{lccc}
\tablecaption{Stellar Parameters for TOI-560 and TOI-5082
\label{tab:stellar_params}}
\tabletypesize{\footnotesize}
\tablehead{
\colhead{Parameter} &
\colhead{TOI-560 (TIC 101011575)} &
\colhead{TOI-5082 (TIC 437011608)} &
\colhead{Source}
}
\startdata
$G_{\rm Gaia}$                                   & $9.28 \pm 0.01$ & $8.09 \pm 0.01$ & Gaia DR3 \citep{Gaia2023} \\
$BP_{\rm Gaia}$                                  & $9.89 \pm 0.01$ & $8.44 \pm 0.01$ & Gaia DR3 \citep{Gaia2023}\\
$RP_{\rm Gaia}$                                  & $8.54 \pm 0.01$ & $7.57 \pm 0.01$ & Gaia DR3 \citep{Gaia2023}\\
Parallax (mas)                                   & $31.657 \pm 0.015$ & $23.238 \pm 0.026$ & Gaia DR3 \citep{Gaia2023}\\ \hline
Effective temperature, $\teff$ (K)               & $4573.8^{+22.3}_{-57.9}$ & $5674.6^{+34.8}_{-39.4}$ & This work \\
Metallicity, [Fe/H] (dex)                        & $-0.08 \pm 0.07$ & $-0.05 \pm 0.14$ & This work \\
Surface gravity, $\log g$ (dex)                  & $4.63 \pm 0.02$ & $4.49 \pm 0.04$ & This work \\
Stellar mass, $M_{\star}$ ($\msun$)              & $0.70 \pm 0.02$ & $0.95 \pm 0.07$ & This work \\
Stellar radius, $R_{\star}$ ($\rsun$)            & $0.67 \pm 0.01$ & $0.92 \pm 0.02$ & This work \\
Stellar luminosity, $L_{\star}$ ($L_{\odot}$)    & $0.18 \pm 0.01$ & $0.79 \pm 0.02$ & This work \\
Stellar density, $\rho_{\star}$ (g\,cm$^{-3}$)   & $2.34 \pm 0.13$ & $1.22 \pm 0.14$ & This work \\
Isochrone-fitted age (Gyr)                       & $4.3^{+3.8}_{-2.7}$ & $3.9^{+4.6}_{-2.6}$ & This work (not used$^{*}$) \\
Distance (pc)                                    & $31.71^{+0.48}_{-0.41}$ & $43.00^{+0.63}_{-0.60}$ & This work \\
\enddata
\tablecomments{
Stellar parameters were derived using the \texttt{isoclassify} software \citep{isoclassify_code}, following the isochrone-based methodology described in \citet{Huber2017,Berger2020,Berger2023}. Reported values correspond to the median of the posterior distributions, with uncertainties representing the central 68\% credible intervals.\\
* The isochrone-fitted age is poorly constrained, and we therefore adopt the age described in Section~\ref{sec:discussion} throughout the paper.
}
\end{deluxetable*}

\section{Data Analysis} \label{sec:analysis}

\subsection{Joint Light Curve and Rossiter--McLaughlin Modeling}

We used a modified version of the {\tt allesfitter}\footnote{\url{https://github.com/wangxianyu7/Allesfitters},\\\url{https://doi.org/10.5281/zenodo.21792490}} package \citep{allesfitter-paper, allesfitter-code, Wang2024}, in which the transit light curves are modeled using the {\tt PyTransit} package \citep{Parviainen2015}, and incorporated the RM model of \citet{Hirano2011}, building on the earlier analytic treatment of \citet{Ohta2005}, to jointly analyze the {\it TESS} photometry and Rossiter--McLaughlin radial velocity observations for TOI-560 and TOI-5082. This approach enables a self-consistent determination of the sky-projected stellar obliquity, $\lambda$, by simultaneously modeling the transit light curves and the spectroscopic RM signal obtained with KPF. For each system, the {\it TESS} data include multiple transits of planet~b observed at a 2-minute cadence across several sectors (see Section~\ref{sec:tess_lc}), while the RM measurements consist of a single, densely sampled transit sequence.

The transit model for each planet is parameterized by the planet-to-star radius ratio, $R_{\rm p}/R_{\star}$, the scaled sum of radii $(R_{\star}+R_{\rm p})/a$, the cosine of the orbital inclination $\cos i_{\rm p}$, the orbital period $P$, and the transit mid-time $T_0$. As no out-of-transit Doppler observations are available for either system, the orbital radial velocity semi-amplitudes ($K_b$) were not included as free parameters, and circular orbits were assumed throughout the analysis.

The spectroscopic time series were modeled as the sum of the RM signal and a low-order polynomial baseline. Specifically, we included an additive quadratic function of time to capture residual orbital motion and stellar variability local to the transit window. Given the limited temporal baseline, this polynomial baseline effectively absorbs slowly varying signals without significantly affecting the recovered RM parameters, particularly $\lambda$, which is primarily constrained by the in-transit anomaly. The RM component introduces two additional parameters: the sky-projected spin--orbit angle, $\lambda$, and the projected stellar rotational velocity, $\vsini$. Limb darkening was described using a quadratic law, parameterized via the transformed coefficients $q_1$ and $q_2$ following \citet{Kipping2013}, with independent sets of coefficients for the {\it TESS} and KPF bandpasses.

Instrumental and astrophysical white-noise contributions were incorporated through free error-scaling terms for each dataset. For the {\it TESS} photometry, we included a multiplicative noise inflation parameter, $\ln \sigma_{\mathrm{{\it TESS}}}$, while an additive jitter term, $\ln \sigma_{\mathrm{jitter;KPF}}$, was added in quadrature to the formal KPF radial velocity uncertainties.

We adopted uniform priors, $\mathcal{U}(a,b)$, for all fitted parameters unless otherwise stated. An informative Gaussian prior was imposed on the stellar mean density, based on the spectroscopic and isochrone-derived stellar properties obtained with {\tt isoclassify}. In modeling the Rossiter--McLaughlin signal using the formalism of \citet{Hirano2011}, we additionally included priors on the stellar microturbulent and macroturbulent velocities, $V_\xi$ and $V_\zeta$. These parameters were assigned Gaussian priors centered on values estimated from empirical relations, with a standard deviation of 1 km\,s$^{-1}$. For $V_\xi$ and $V_\zeta$, we adopted the calibrations from \citet{Bruntt2010} and \citet{Doyle2014}, respectively, when the stellar parameters fell within their stated validity ranges. Outside these ranges, we instead employed the empirical relations calibrated by the Gaia--ESO Survey working groups \citep{Blanco2014HighResolutionSpectralLibrary, Blanco2014iSpec, Blanco2019iSpec}.

Posterior sampling was performed using the dynamic nested sampling algorithm implemented in the {\tt dynesty} package \citep{Higson2019,DYNESTY} through {\tt allesfitter}. We adopted an initial set of 500 live points, a random-walk ({\tt rwalk}) proposal scheme, and a stopping criterion of $\Delta \ln Z < 0.01$, where $Z$ denotes the Bayesian evidence. Unlike traditional Markov Chain Monte Carlo methods, dynamic nested sampling does not employ a fixed number of walkers or chain steps. The final posterior distributions were constructed from approximately $5.0\times10^{4}$ and $3.4\times10^{4}$ posterior samples for TOI-560 and TOI-5082, respectively. Convergence was assessed by verifying that the sampler reached the adopted evidence tolerance and that the posterior distributions were well sampled upon termination. We additionally examined the residual diagnostics produced by {\tt allesfitter}, including Anderson–Darling, Augmented Dickey–Fuller, Durbin–Watson, and Ljung–Box statistics, to evaluate the adequacy of the final fits.

\begin{figure*}[htbp]
    \centering
    \includegraphics[width=0.32\textwidth]{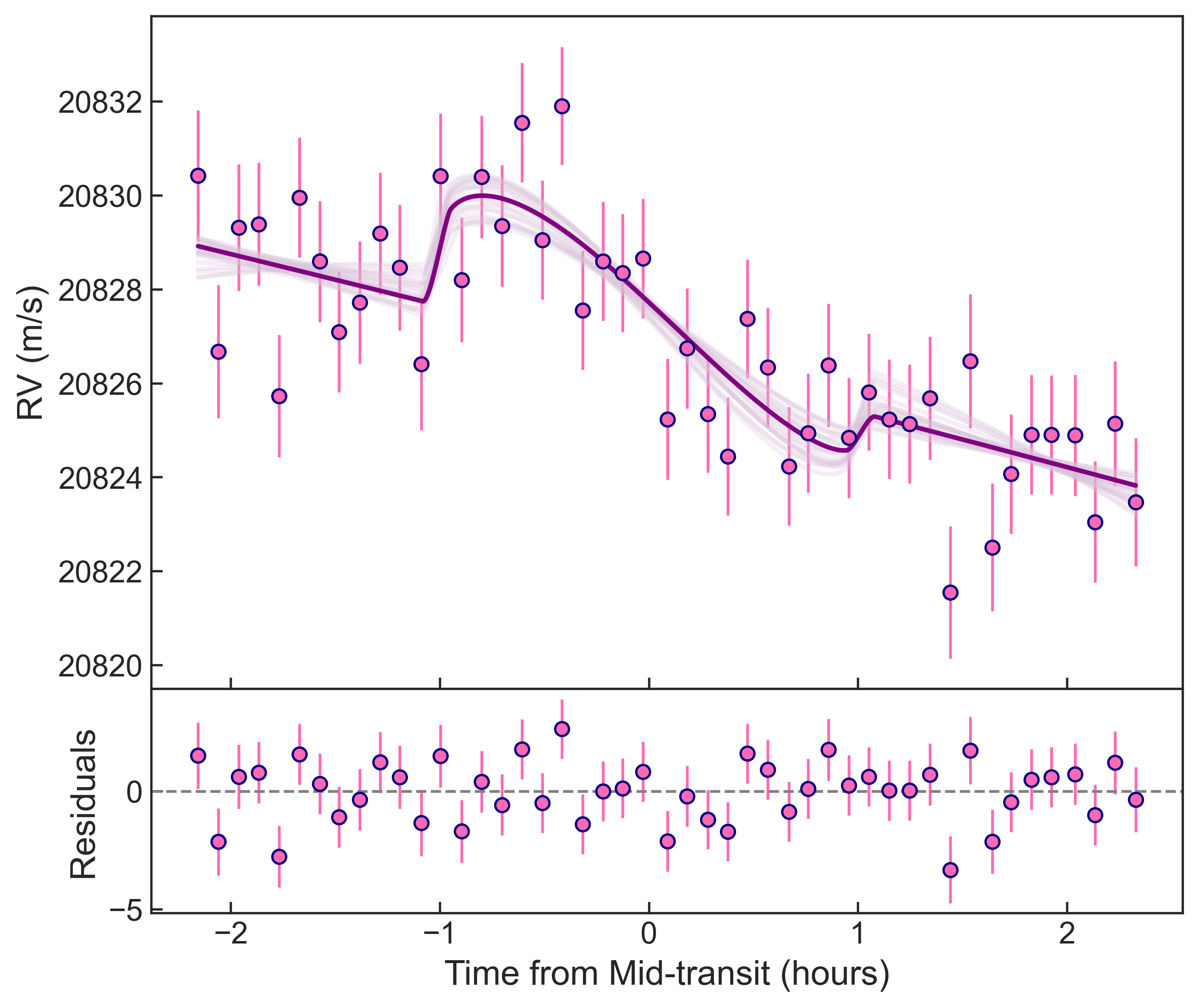}
    \includegraphics[width=0.3\textwidth]{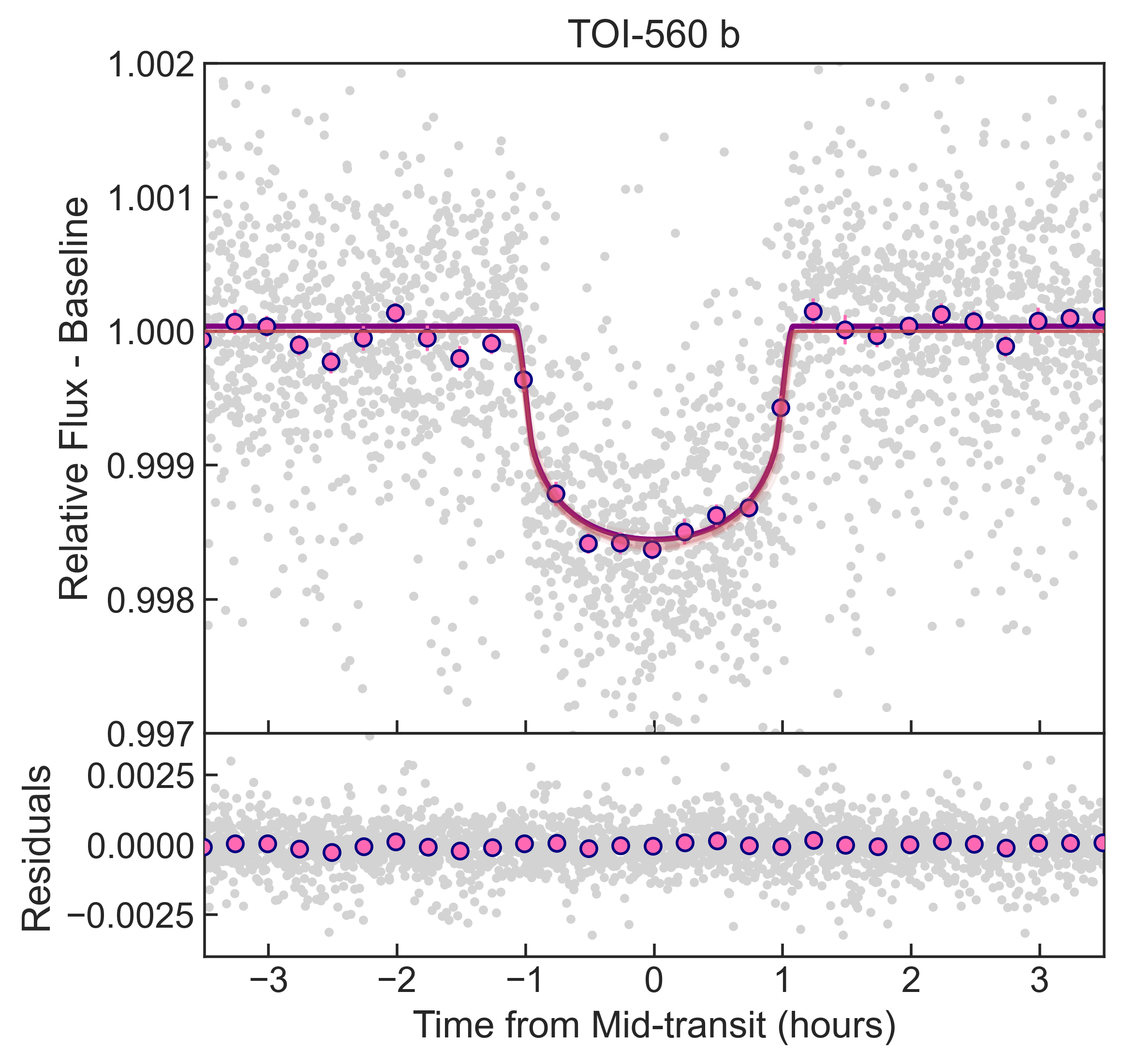}
    \includegraphics[width=0.292\textwidth]{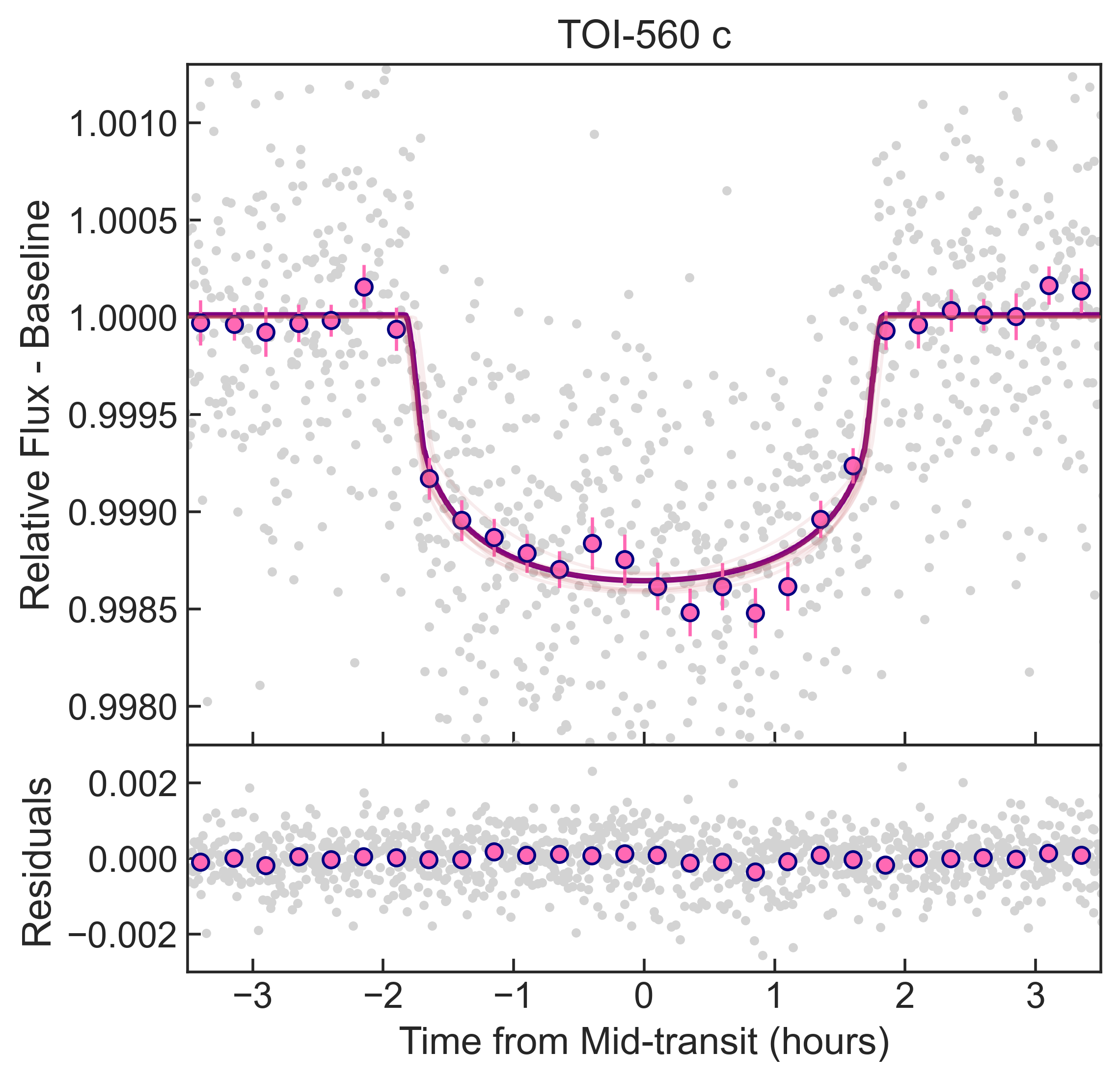}
    \caption{\normalsize Left: Radial velocity measurements obtained with KPF during the transit of TOI-560~b on UTC 2024 December 16. The top panel shows the observed velocities together with model realizations drawn from the posterior distribution (light pink curves); the dark purple curve denotes the best-fitting model, corresponding to a sky-projected obliquity of $\lambda_b = -25 \pm 16^{\circ}$. The lower panel displays the residuals after subtracting the best-fit model. 
    Middle and Right: Phase-folded, normalized {\it TESS} light curves of TOI-560~b (middle) and TOI-560~c (right), shown as a function of time from mid-transit. The data were obtained at a 2-minute cadence. Pink points with blue outlines represent the binned photometry, while the light red curves indicate 20 random posterior samples. The dark purple curve shows the median transit model from {\tt allesfitter}. Residuals relative to the best-fit model are shown in the lower panels.}
\label{Fig:TOI560_alles}
\end{figure*}

\begin{figure*}[htbp]
    \centering
    \includegraphics[width=0.35\textwidth]{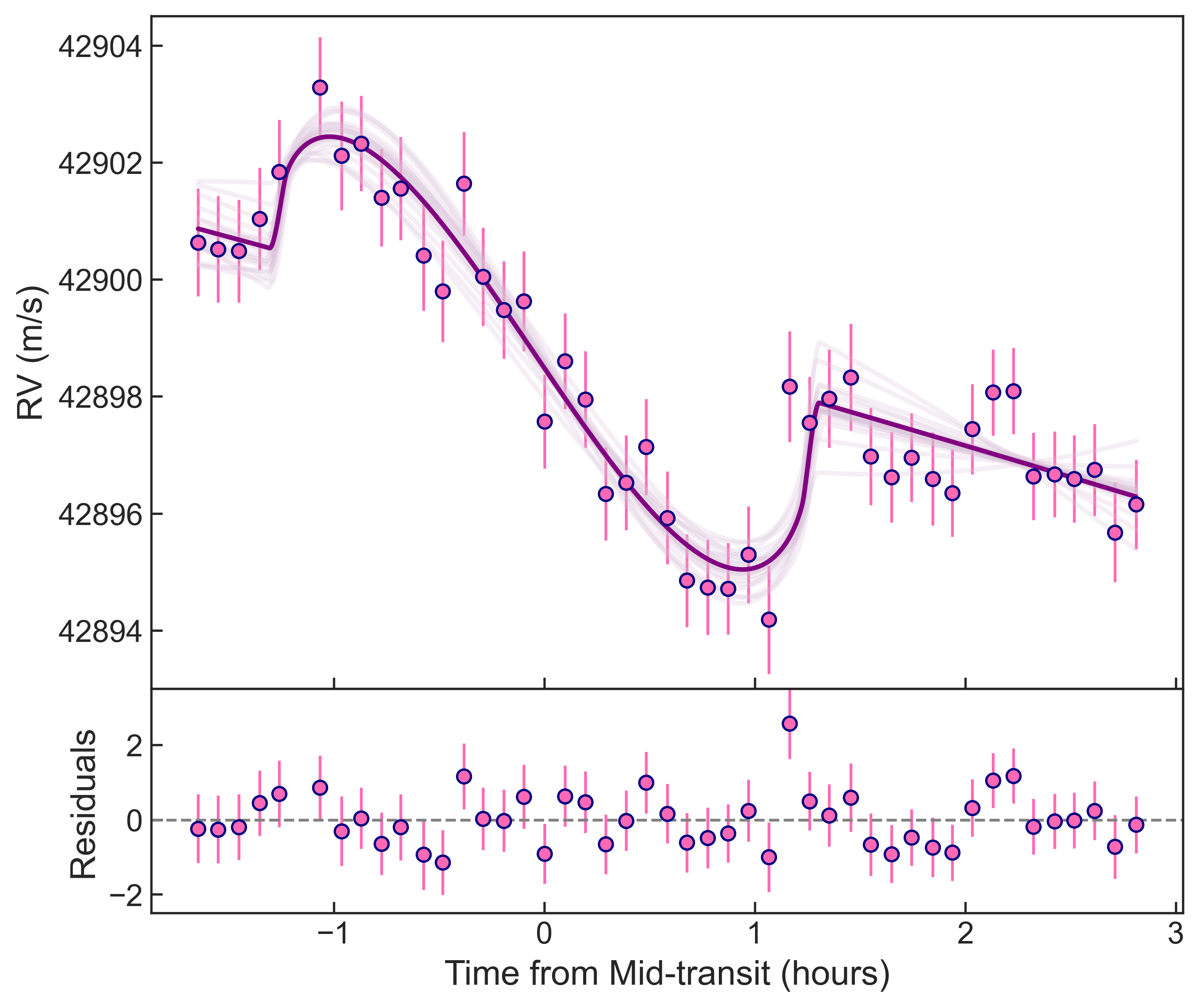}
    \includegraphics[width=0.32\textwidth]{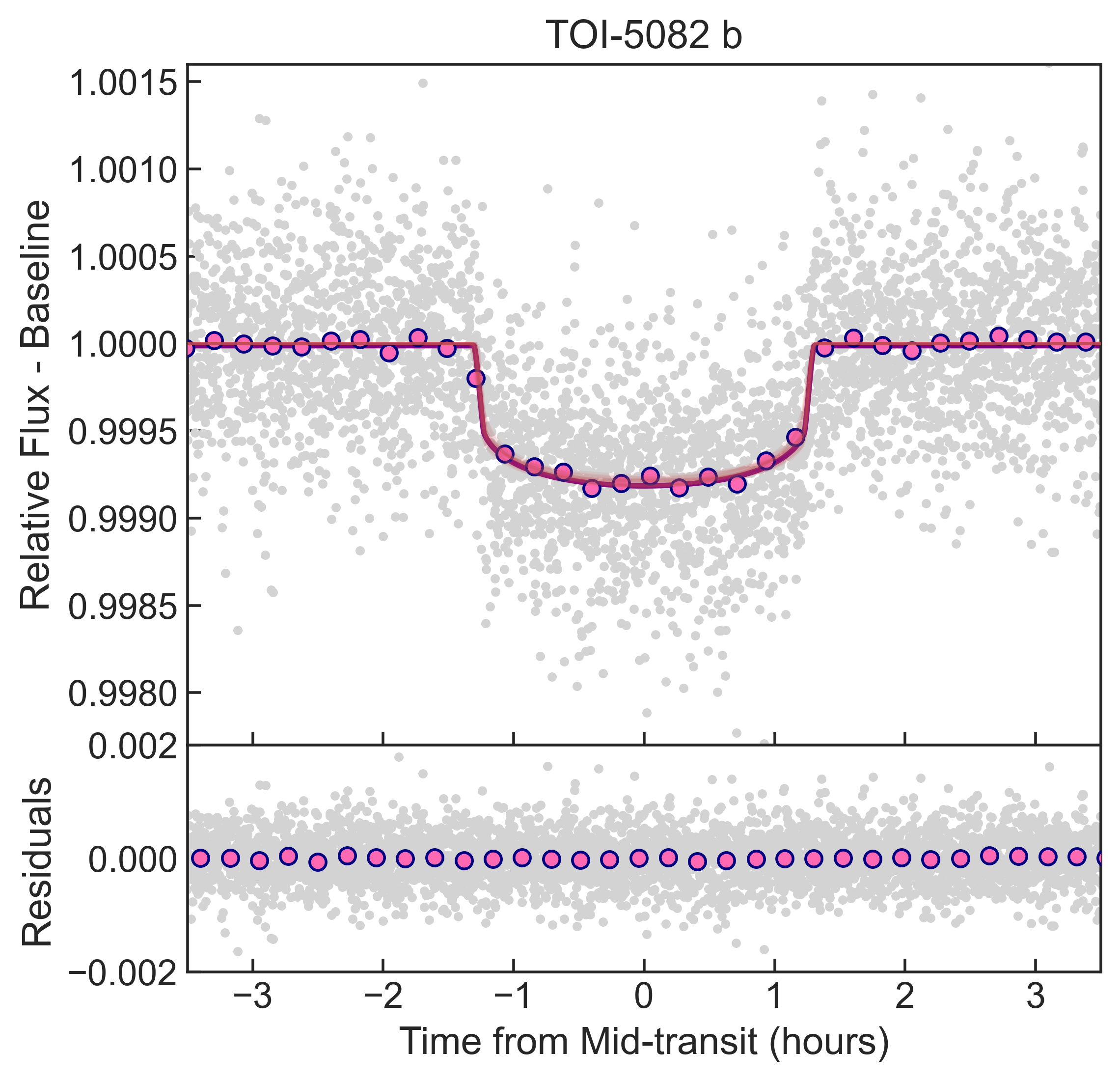}
    \caption{\normalsize Joint RM and transit fit for TOI-5082~b, following the plotting conventions of Figure~\ref{Fig:TOI560_alles}. The KPF observations were obtained on UTC 2024 November 26, and the best-fitting RM model corresponds to a sky-projected obliquity of $\lambda_b = 19^{+17}_{-13}$\,deg.}
\label{Fig:TOI5082_alles}
\end{figure*}

The best-fitting RM models and detrended {\it TESS} transit light curves for TOI-560~b and TOI-5082~b are shown in Figures~\ref{Fig:TOI560_alles} and \ref{Fig:TOI5082_alles}, respectively. The inferred median values and corresponding 68\% credible intervals for the fitted parameters, together with the derived physical quantities computed from the joint posterior distributions and stellar properties, are summarized in Tables~\ref{tab:toi560_all_params} and \ref{tab:toi5082_all_params}.

\begin{deluxetable*}{@{}cccc@{}}
\tablecaption{Joint {\tt allesfitter} Model Parameters for TOI-560
\label{tab:toi560_all_params}}
\tabletypesize{\footnotesize}
\tablehead{
\colhead{Parameter} &
\colhead{Prior} &
\colhead{Value} &
\colhead{Unit}
}
\startdata
\multicolumn{4}{c}{\textbf{Fitted Parameters}} \\ \hline
$R_b / R_\star$ & $\mathcal{U}(0,0.1)$ & $0.03798_{-0.00064}^{+0.00069}$ &  \\ 
$(R_\star + R_b) / a_b$ & $\mathcal{U}(0,0.5)$ & $0.05612_{-0.00095}^{+0.0010}$ &  \\ 
$\cos i_b$ & $\mathcal{U}(0,0.2)$ & $0.0346\pm0.0016$ &  \\ 
$T_{0;b}-2457000^{\ast}$ & $\mathcal{U}(2611.65, 2611.85)$ & $2611.75734\pm0.00044$ & BJD \\ 
$P_b$ & $\mathcal{U}(6.23, 6.50)$ & $6.3980492\pm0.0000032$ & d \\ 
$\sqrt{e_b}\cos\omega_b$ & fixed & 0.0 &  \\ 
$\sqrt{e_b}\sin\omega_b$ & fixed & 0.0 &  \\ \hline
$R_c / R_\star$ & $\mathcal{U}(0,0.1)$ & $0.03370\pm0.00067$ &  \\ 
$(R_\star + R_c) / a_c$ & $\mathcal{U}(0,0.5)$ & $0.02719_{-0.00048}^{+0.00051}$ &  \\ 
$\cos i_c$ & $\mathcal{U}(0,0.2)$ & $0.0098_{-0.0017}^{+0.0015}$ &  \\ 
$T_{0;c}-2457000^{\ast}$ & $\mathcal{U}(2628.55, 2628.75)$ & $2628.63566\pm0.00088$ & BJD \\ 
$P_c$ & $\mathcal{U}(18.78, 18.98)$ & $18.879242_{-0.000031}^{+0.000028}$ & d \\ 
$\sqrt{e_c}\cos\omega_c$ & fixed & 0.0 &  \\ 
$\sqrt{e_c}\sin\omega_c$ & fixed & 0.0 &  \\ \hline
$q_{1;\mathrm{TESS}}$ & $\mathcal{U}(0,1)$ & $0.61_{-0.20}^{+0.23}$ &  \\ 
$q_{2;\mathrm{TESS}}$ & $\mathcal{U}(0,1)$ & $0.25_{-0.16}^{+0.25}$ &  \\ 
$q_{1;\mathrm{KPF}}$ & $\mathcal{U}(0,1)$ & $0.58_{-0.35}^{+0.29}$ &  \\ 
$q_{2;\mathrm{KPF}}$ & $\mathcal{U}(0,1)$ & $0.56_{-0.36}^{+0.31}$ &  \\ 
$\ln\sigma_{\mathrm{TESS}}$ & $\mathcal{U}(-10,0)$ & $-7.178\pm0.011$ & ln(rel.\ flux) \\ 
$\ln\sigma_{\mathrm{jitter;KPF}}$ & $\mathcal{U}(-10,0)$ & $-6.82\pm0.16$ & ln(km\,s$^{-1}$) \\ 
$\lambda_\star$ & $\mathcal{U}(-180,180)$ & $-25\pm16$ & deg \\ 
$\vsini$ & $\mathcal{U}(0,10)$ & $2.76_{-0.63}^{+0.67}$ & km\,s$^{-1}$ \\ 
$V_\xi$ & $\mathcal{N}(0.87,1)$ & $1.09_{-0.69}^{+0.89}$ & km\,s$^{-1}$ \\ 
$V_\zeta$ & $\mathcal{N}(3.7,1)$ & $3.74\pm0.98$ & km\,s$^{-1}$ \\ \hline
\multicolumn{4}{c}{\textbf{Derived Parameters}} \\ \hline
$a_b/R_\star$ & -- & $18.49\pm0.32$ &  \\ 
$R_b$ & -- & $2.900\pm0.067$ & $R_\oplus$ \\ 
$a_b$ & -- & $0.0602\pm0.0013$ & AU \\ 
$i_b$ & -- & $88.015\pm0.095$ & deg \\ 
$b_{\rm tra;b}$ & -- & $0.641_{-0.022}^{+0.021}$ &  \\ 
$T_{\rm tot;b}$ & -- & $2.160_{-0.028}^{+0.030}$ & h \\ 
$T_{\rm full;b}$ & -- & $1.898\pm0.037$ & h \\ 
$T_{\rm eq;b}$ & -- & $679\pm18$ & K \\ 
$\delta_{\rm tr;b}$ & -- & $1.588\pm0.041$ & ppt \\ \hline
$a_c/R_\star$ & -- & $38.01\pm0.69$ &  \\ 
$R_c$ & -- & $2.572\pm0.063$ & $R_\oplus$ \\ 
$a_c$ & -- & $0.1237\pm0.0029$ & AU \\ 
$i_c$ & -- & $89.438_{-0.087}^{+0.096}$ & deg \\ 
$b_{\rm tra;c}$ & -- & $0.373_{-0.059}^{+0.050}$ &  \\ 
$T_{\rm tot;c}$ & -- & $3.656_{-0.045}^{+0.049}$ & h \\ 
$T_{\rm full;c}$ & -- & $3.380_{-0.053}^{+0.056}$ & h \\ 
$T_{\rm eq;c}$ & -- & $474\pm12$ & K \\ 
$\delta_{\rm tr;c}$ & -- & $1.372_{-0.048}^{+0.051}$ & ppt \\ \hline
$P_c/P_b$ & -- & $2.9507809_{-0.0000051}^{+0.0000048}$ &  \\ 
$\rho_{\star;\rm combined}$ & -- & $2.92\pm0.16$ & g\,cm$^{-3}$ \\
\enddata
\tablecomments{
$\mathcal{U}(a,b)$ denotes a uniform prior between $a$ and $b$. \\
Gaussian priors with $\sigma=1$ km\,s$^{-1}$ were adopted for the micro- and macroturbulent velocities. \\
$^{\ast}${\tt allesfitter} will automatically shift the input prior epoch into the data center. For TOI-560~b, the epoch is shifted by 172 periods, and 21 periods for TOI-560~c.}
\end{deluxetable*}

\begin{deluxetable*}{@{}cccc@{}}
\tablecaption{Joint {\tt allesfitter} Model Parameters for TOI-5082
\label{tab:toi5082_all_params}}
\tabletypesize{\footnotesize}
\tablehead{
\colhead{Parameter} &
\colhead{Prior} &
\colhead{Value} &
\colhead{Unit}
}
\startdata
\multicolumn{4}{c}{\textbf{Fitted Parameters}} \\ \hline
$R_b / R_\star$ & $\mathcal{U}(0,0.1)$ & $0.02650_{-0.00046}^{+0.00051}$ &  \\ 
$(R_\star + R_b) / a_b$ & $\mathcal{U}(0,0.5)$ & $0.0861_{-0.0044}^{+0.0073}$ &  \\ 
$\cos i_b$ & $\mathcal{U}(0,0.2)$ & $0.031\pm0.016$ &  \\ 
$T_{0;b}-2457000^{\ast}$ & $\mathcal{U}(3072.65, 3072.85)$ & $3072.78493_{-0.00034}^{+0.00031}$ & BJD \\ 
$P_b$ & $\mathcal{U}(4.14, 4.34)$ & $4.2403486\pm0.0000033$ & d \\ 
$\sqrt{e_b}\cos\omega_b$ & fixed & 0.0 &  \\ 
$\sqrt{e_b}\sin\omega_b$ & fixed & 0.0 &  \\ \hline
$q_{1;\mathrm{TESS}}$ & $\mathcal{U}(0,1)$ & $0.29_{-0.11}^{+0.19}$ &  \\ 
$q_{2;\mathrm{TESS}}$ & $\mathcal{U}(0,1)$ & $0.32_{-0.21}^{+0.30}$ &  \\ 
$q_{1;\mathrm{KPF}}$ & $\mathcal{U}(0,1)$ & $0.80_{-0.22}^{+0.14}$ &  \\ 
$q_{2;\mathrm{KPF}}$ & $\mathcal{U}(0,1)$ & $0.58_{-0.32}^{+0.27}$ &  \\ 
$\ln\sigma_{\mathrm{TESS}}$ & $\mathcal{U}(-10,0)$ & $-7.783\pm0.011$ & ln(rel.\ flux) \\ 
$\ln\sigma_{\mathrm{jitter;KPF}}$ & $\mathcal{U}(-10,0)$ & $-9.15_{-0.59}^{+0.73}$ & ln(km\,s$^{-1}$) \\ 
$\lambda_\star$ & $\mathcal{U}(-180,180)$ & $19_{-13}^{+17}$ & deg \\ 
$\vsini$ & $\mathcal{U}(0,10)$ & $6.41_{-0.93}^{+1.10}$ & km\,s$^{-1}$ \\ 
$V_\xi$ & $\mathcal{N}(1.0,1)$ & $1.16_{-0.73}^{+0.92}$ & km\,s$^{-1}$ \\ 
$V_\zeta$ & $\mathcal{N}(2.86,1)$ & $2.83\pm1.00$ & km\,s$^{-1}$ \\ \hline
\multicolumn{4}{c}{\textbf{Derived Parameters}} \\ \hline
$a_b/R_\star$ & -- & $11.92_{-0.93}^{+0.64}$ &  \\ 
$R_b$ & -- & $2.69\pm0.13$ & $R_\oplus$ \\ 
$a_b$ & -- & $0.0513_{-0.0044}^{+0.0037}$ & AU \\ 
$i_b$ & -- & $88.23\pm0.93$ & deg \\ 
$b_{\rm tra;b}$ & -- & $0.37_{-0.18}^{+0.15}$ &  \\ 
$T_{\rm tot;b}$ & -- & $2.609_{-0.023}^{+0.028}$ & h \\ 
$T_{\rm full;b}$ & -- & $2.447_{-0.025}^{+0.030}$ & h \\ 
$T_{\rm eq;b}$ & -- & $1065_{-34}^{+47}$ & K \\ 
$\delta_{\rm tr;b}$ & -- & $0.802\pm0.019$ & ppt \\ \hline
$\rho_{\star;\rm combined}$ & -- & $1.78_{-0.39}^{+0.30}$ & g\,cm$^{-3}$ \\
\enddata
\tablecomments{
$\mathcal{U}(a,b)$ denotes a uniform prior between $a$ and $b$. \\
Gaussian priors with $\sigma=1$ km\,s$^{-1}$ were adopted for the micro- and macroturbulent velocities. \\
$^{\ast}${\tt allesfitter} will automatically shift the input prior epoch into the data center. For TOI-5082~b, the epoch is shifted by 50 periods.}
\end{deluxetable*}

\subsection{Stellar Rotation Analysis}
\label{sec:star_Prot}

We investigate stellar rotation using long-baseline photometric observations from \textit{TESS}. Stellar rotation periods are inferred from quasi-periodic brightness modulations induced by starspots rotating in and out of view.

For TOI-5082, we perform our own rotation analysis using \textit{TESS} SPOC light curves. The data are first preprocessed by removing NaN values and applying iterative sigma-clipping to eliminate outliers. The light curves are then normalized and analyzed on a sector-by-sector basis to account for spot evolution and changes in modulation amplitude. The cleaned sectors are subsequently combined to maximize the time baseline for period determination.

We compute a Lomb--Scargle periodogram \citep{Lomb1976,Scargle1982,VanderPlas2018} over a broad range of trial periods relevant for stellar rotation. The dominant peak is identified and modeled with a Gaussian profile, from which we estimate the rotation period and its associated uncertainty. This approach accounts for the finite width of the periodogram peak arising from spot evolution and possible differential rotation.

For TOI-5082, we detect a clear rotational modulation with a period of $P_{\rm rot} = 8.93 \pm 0.54$ days. This period is well separated from the planetary orbital period, indicating that the system is not tidally synchronized. Using the relation $v_{\rm eq} = 2\pi R_{\star}/P_{\rm rot}$, we derive an equatorial rotation velocity of $v_{\rm eq} = 5.21 \pm 0.33~\mathrm{km\,s^{-1}}$. We further estimate a stellar age of $734.13^{+108.33}_{-96.4}$ Myr using {\tt gyro-interp} \citep{Bouma2023}.

For TOI-560, we do not rely on our own periodogram analysis, but instead adopt the stellar rotation period of $P_{\rm rot} = 12.2 \pm 0.1$ days from \citet{ElMufti2023}. In that work, the rotation period was derived from Lomb--Scargle analyses of multi-season SuperWASP photometry \citep{Pollacco2006,Butters2010}, which show a consistent $\sim$12 day periodicity across four observing seasons. The resulting periodogram peaks are statistically significant in each season, and the average period yields $P_{\rm rot} = 12.2 \pm 0.1$ days. This value is further supported by independent activity diagnostics, including chromospheric emission indicators, empirical rotation--activity relations, and consistency with spectroscopic constraints on $\vsini$ \citep{ElMufti2023}.

For completeness, we performed an independent Lomb--Scargle analysis \citep{Lomb1976,Scargle1982,VanderPlas2018} of the \textit{TESS} light curves for TOI-560, which yielded a periodicity of $P_{\rm rot} = 5.82 \pm 0.53$ days. This signal lies close to the orbital period of planet~b ($P_{\rm orb,b} \approx 6.4$ days). As discussed by \citet{ElMufti2023}, the \textit{TESS} light curves exhibit significant power at periods of $\sim$5--7 days, corresponding to approximately half of the true rotation period. This behavior is attributed to a combination of the first harmonic of the rotation signal ($P_{\rm rot}/2$), evolving starspot distributions, and the effects of the presearch data conditioning (PDC) pipeline. Additionally, the limited time baseline of individual \textit{TESS} sectors ($\lesssim 27$ days) reduces sensitivity to longer rotation periods near $\sim$12 days, further biasing the periodogram toward shorter harmonic signals. We therefore interpret the $\sim$5.8 day signal as a harmonic of the true rotation period and adopt $P_{\rm rot} = 12.2 \pm 0.1$ days throughout this work. Under this assumption, we derive an equatorial rotation velocity of $v_{\rm eq} = 2.78 \pm 0.05~\mathrm{km\,s^{-1}}$, and estimate a stellar age of $1.14^{+0.15}_{-0.27}$~Gyr using {\tt gyro-interp} \citep{Bouma2023}.

\paragraph{Stellar Age Assessment.}
The isochronal age obtained using \texttt{isoclassify} in Section~\ref{sec:stellar_para} provides weak constraints on stellar age due to degeneracies among stellar parameters and the weak age-dependence of observable properties of main-sequence stars. The gyrochronology age derived from the stellar rotation period in Section~\ref{sec:star_Prot} is more informative, but remains subject to considerable uncertainties related to calibration, magnetic activity evolution, and possible deviations from standard spin-down relations. We therefore explored alternative age diagnostics for both systems.

For TOI-5082, \citet{Gagne2026} identify the host star (TIC~437011608) as a candidate member of the CRIUS197 stellar association with a BANYAN~$\Sigma$ membership probability of 92.3\%. CRIUS197 was originally identified by \citet{Moranta2022} as a nearby candidate moving group, although it was not included among the subset of groups studied in detail due to its large spatial extent in the Galactic $Z$ direction. \citet{Robbins2023} reported a preliminary gyrochronology-based age of $180 \pm 9$~Myr for CRIUS197, while noting that two candidate members are clear outliers in the rotation period--color sequence, suggesting that the group may include interlopers or represent a heterogeneous population. We therefore adopt $180 \pm 9$~Myr as a tentative age estimate for TOI-5082, with the caveat that the age of CRIUS197 remains uncertain. For TOI-560, we adopt an age of $480 \pm 190$~Myr from \citet{Barragan2022_TOI560}, derived from the chromospheric activity index $\log R'_{\rm HK}$ using the empirical activity--age relation of \citet{Mamajek2008}.

We emphasize that the gyrochronological ages derived above are reported for completeness but are not adopted in our subsequent analysis.

\subsection{True Obliquity}
\subsubsection{Projected stellar rotation velocity from CCF modeling}

In addition to the equatorial rotation velocities inferred from the stellar rotation periods and radii in Section~\ref{sec:star_Prot}, we obtained spectroscopic constraints on $\vsini$ for TOI-5082 and TOI-560 from the widths of the cross-correlation functions (CCFs) in the KPF Level~2 data products. The CCF width provides an independent spectroscopic estimate of the projected stellar rotation velocity, $\vsini$, through rotational broadening of the composite line profile. This spectroscopic constraint is combined with the rotation-based estimate of the equatorial velocity in the joint inference of the stellar inclination. The analysis was restricted to out-of-transit exposures to avoid contamination from the Rossiter--McLaughlin effect, and exposures with anomalously low flux were excluded.

For each target, the order-by-order CCFs were shifted into the stellar rest frame using the radial velocities reported in the KPF headers and combined into a weighted master CCF. The master CCF was modeled as an absorption profile consisting of a low-order polynomial baseline and a broadened line profile. The broadening kernel was constructed from a rotational profile following the formalism of \citet{Gray2005}, together with instrumental and turbulent broadening terms. The rotational kernel adopted a linear limb-darkening coefficient derived from the transit analysis.

Rather than approximating the KPF instrumental profile as a simple Gaussian corresponding to the nominal resolving power, we adopted an empirical instrumental kernel calibrated from laser-frequency-comb (LFC) cross-correlation functions. Following fits to the LFC profile, the instrumental line-spread function was represented as the convolution of a Gaussian and a top-hat function, with $\sigma_{\rm instr}=0.68~{\rm km/s}$ and top-hat width $w_{\rm instr}=3.12~{\rm km/s}$. The stellar CCF was modeled as the convolution of the rotational broadening profile, the empirical instrumental profile, a Gaussian macroturbulent broadening component, and an additional Lorentzian term that accounts for residual wing structure not captured by the rotational and instrumental broadening alone. The resulting broadening kernel was normalized to unit peak amplitude to ensure a stable separation between the CCF depth and width.

The projected stellar rotation velocity was determined via non-linear least-squares fitting of the master CCF. Macroturbulent broadening was treated as a free parameter with a Gaussian prior centered on the value predicted by the empirical relations of \citet{Doyle2014}, thereby accounting for covariance between rotational and turbulent broadening. Formal parameter uncertainties were estimated from the covariance matrix of the best-fit solution. To account for order-to-order and exposure-level systematics, we additionally performed a bootstrap analysis in which exposures and spectral orders were resampled with replacement. In each bootstrap realization, the center of the macroturbulence prior was perturbed according to its adopted uncertainty, and the final uncertainty on $\vsini$ was taken from the resulting bootstrap distribution.

Because the CCF is a mask-weighted composite profile rather than an individual stellar absorption line, its width can also depend on mismatch between the numerical mask and the stellar spectrum and on the line-weighting scheme used to construct the CCF \citep[e.g.,][]{Rainer2023}. Our order-level bootstrap captures some sensitivity to the set of lines contributing to the CCF, but it may not capture mask-related systematic effects common to all orders. The reported uncertainties should therefore be interpreted as statistical and order-to-order uncertainties under the adopted CCF model. Direct modeling of the extracted stellar spectra would be required for a more rigorous spectroscopic determination of $\vsini$.

We obtain CCF-based estimates of $\vsini = 5.57 \pm 0.29~\mathrm{km\,s^{-1}}$ for TOI-5082 and $\vsini = 2.57 \pm 1.17~\mathrm{km\,s^{-1}}$ for TOI-560. These estimates are consistent within their uncertainties with the equatorial rotation velocities inferred from the stellar rotation analysis in Section~\ref{sec:star_Prot} and with the independent constraints from the Rossiter--McLaughlin modeling: $6.41^{+1.10}_{-0.93}~\mathrm{km\,s^{-1}}$ and $2.76^{+0.67}_{-0.63}~\mathrm{km\,s^{-1}}$ for TOI-5082 and TOI-560, respectively. The agreement with the Rossiter--McLaughlin results provides a useful consistency check, although the quoted CCF-based uncertainties do not include all possible mask-related systematics.

\subsubsection{Inference of the True Obliquity}

The true three-dimensional spin--orbit obliquity, $\psi$, is the angle between the stellar spin axis and the planet's orbital angular momentum vector. While the Rossiter--McLaughlin effect constrains the sky-projected obliquity $\lambda$, recovering $\psi$ also requires the stellar inclination $i_\star$ and orbital inclination $i_p$.

We inferred $i_\star$ using independent constraints on the stellar equatorial rotation velocity $v$ and the projected rotation velocity $\vsini$, following \citet{MasudaWinn2020}. The stellar rotation period and radius constrain $v$, while the spectroscopic CCF analysis provides an independent measurement of $\vsini$. We adopted an isotropic prior on the stellar spin orientation and marginalized over the unknown equatorial velocity following \citet{MasudaWinn2020}, thereby fully propagating the uncertainties in $v$ and $\vsini$ without relying on a naive ratio of the two quantities.

Because spectroscopic measurements of $v\sin i_\star$ are insensitive to the sign of $\cos i_\star$, the inferred posterior has an intrinsic degeneracy between $i_\star$ and $180^\circ-i_\star$. In our Monte Carlo calculation, we accounted for this by randomly assigning the sign of $\cos i_\star$ with equal probability. We then combined samples from the resulting $\cos i_\star$ posterior with paired samples of $\lambda$ and $i_p$ from the joint \texttt{allesfitter} nested-sampling posterior, thereby preserving the covariance between the fitted orbital parameters. For each combined posterior draw, we calculated $\psi$ using
\begin{equation}
\cos\psi =
\sin i_\star\sin i_p\cos\lambda+
\cos i_\star\cos i_p,
\end{equation}
fully propagating the non-Gaussian and geometric uncertainties. The inferred stellar inclinations are broadly consistent with both host stars being viewed at relatively high inclination, with folded posterior medians of $i_\star=64.1^{+17.8}_{-22.5}\arcdeg$ for TOI-560 and $i_\star=80.4^{+6.6}_{-9.1}\arcdeg$ for TOI-5082.

\begin{figure*}
    \centering
    \includegraphics[width=0.49\textwidth]{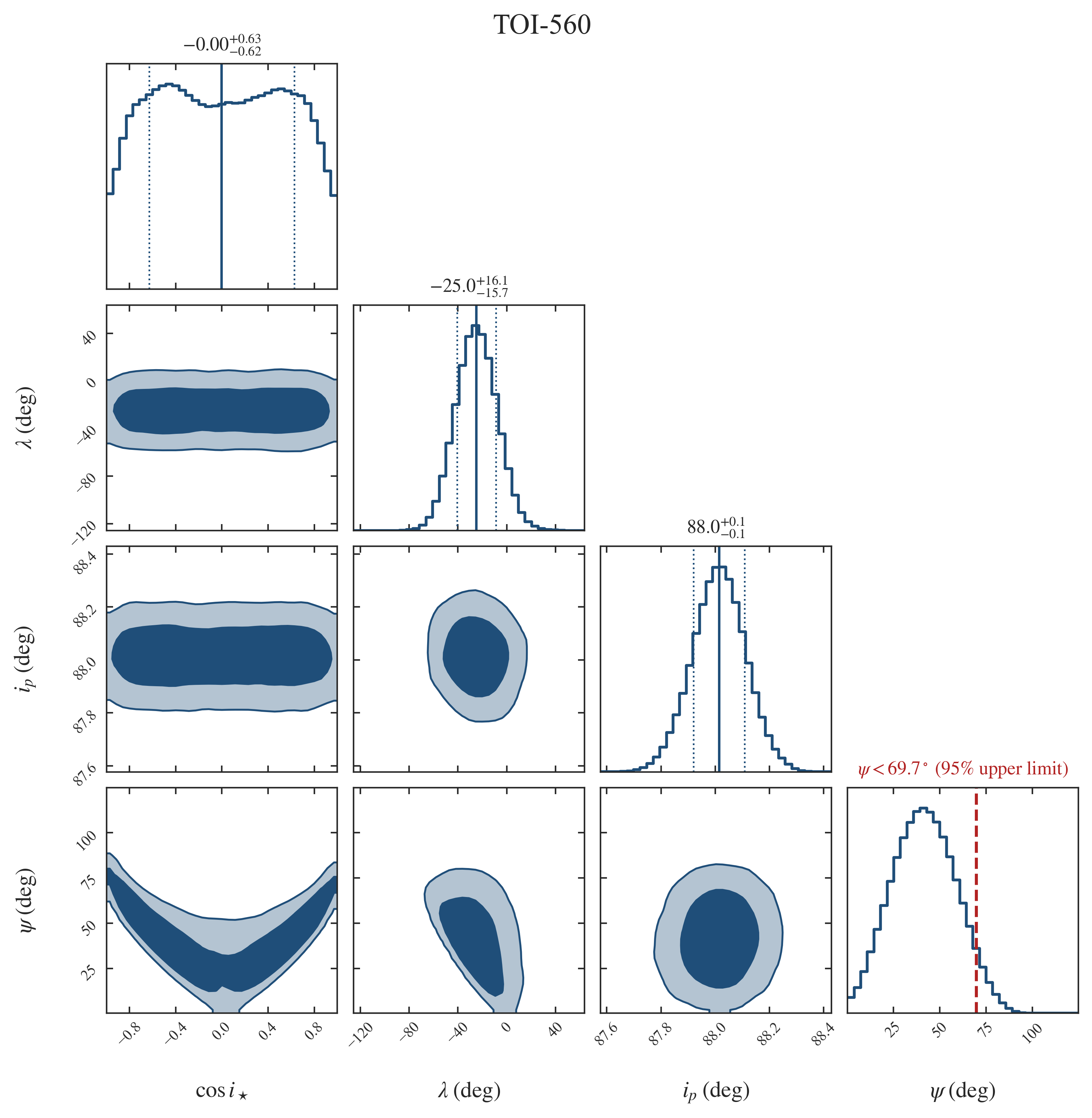}
    \hfill
    \includegraphics[width=0.49\textwidth]{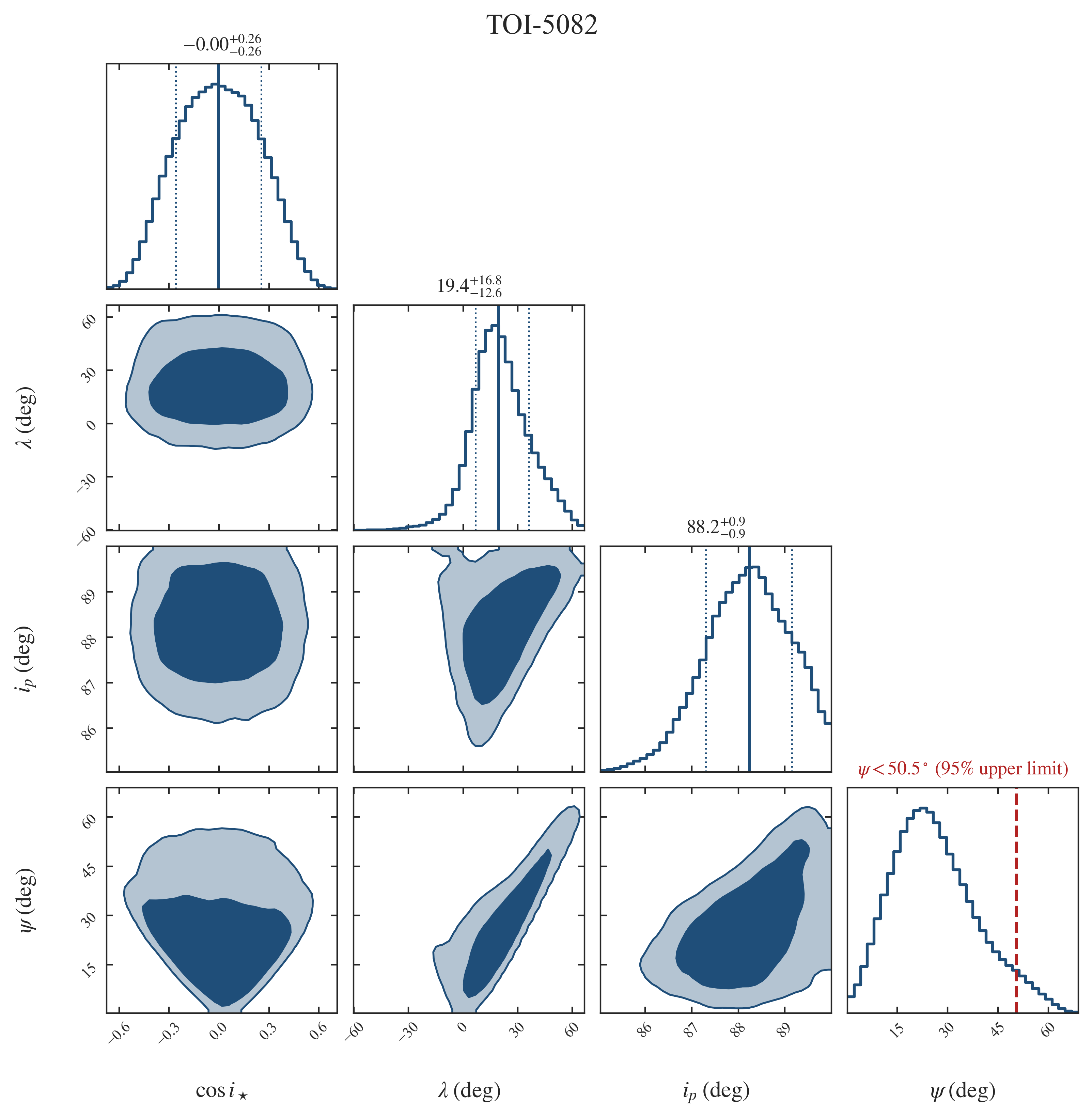}
    \caption{Marginalized and joint posterior distributions of $\cos i_\star$, the sky-projected obliquity $\lambda$, the orbital inclination $i_p$, and the true three-dimensional obliquity $\psi$ for TOI-560\,b (left) and TOI-5082\,b (right). The $\lambda$ and $i_p$ samples are drawn from the joint \texttt{allesfitter} nested-sampling posterior, preserving their covariance, while the $\cos i_\star$ posterior is inferred independently from the stellar radius, rotation period, and projected rotation velocity. The degeneracy between $i_\star$ and $180^\circ-i_\star$ is incorporated by assigning equal probability to the two signs of $\cos i_\star$. The dark and light shaded contours enclose 68\% and 95\% of the two-dimensional posterior probability, respectively. The red dashed lines mark the one-sided 95\% credible upper limits, $\psi<69.7^\circ$ for TOI-560\,b and $\psi<50.5^\circ$ for TOI-5082\,b.}
    \label{fig:true_obliquity_corner}
\end{figure*}

Rather than over-interpreting the median values of $\psi$, we emphasize that the inferred posteriors are broad and reflect significant geometric degeneracies. We therefore report upper limits on the true obliquity. At 95\% posterior probability, we find the one-sided credible upper limits $\psi < 69.7^\circ$ for TOI-560\,b and $\psi < 50.5^\circ$ for TOI-5082\,b. These constraints indicate that both systems are consistent with low-to-moderate true obliquities and show no evidence for strong spin--orbit misalignment. The marginalized and joint posterior distributions of $\cos i_\star$, $\lambda$, $i_p$, and $\psi$ are shown for both systems in Figure~\ref{fig:true_obliquity_corner}.

\section{Discussion: Stellar Obliquity of Young Neptunes} \label{sec:discussion}

Given the uncertainties on the true obliquities, we cannot rule out modest non-zero spin–orbit misalignments in the TOI-560 and TOI-5082 systems. Recently, \citet{Biddle2025Nature} found that the majority of young star--disk systems are consistent with alignment, but exhibit a broad distribution of non-zero obliquities with a characteristic peak at $\sim10^\circ$--$25^\circ$ and possibly higher values. In addition, recent high-resolution ALMA studies have shown that protoplanetary disks themselves are not strictly planar, but can exhibit coherent radial warps at the $\sim$degree level \citep{Winter2025}. One may expect the planetary systems formed in such misaligned disks to be also misaligned with their host star. \citet{Spalding} pointed out that, depending on the relative strength of nodal precession caused by the host star's quadrupole moment and the secular precession caused by the planets themselves,  Kepler-like compact multi-planets systems may precess differentially around the stellar spin axis (if quadrupole dominates), or precess as a group if planets secularly coupled with each other more strongly. In the latter case, mutual inclination between the planets is not expected even though the planets may have a non-zero obliquity with the stellar spin axis \citep[see also][]{Teng,Zhang2025}. TOI-560 and TOI-5082 are relatively young systems, but both stars are currently slow rotators and therefore are expected to possess weak stellar quadrupole moments. As a result, secular planet--planet interactions are likely to dominate over torques from the stellar quadrupole moment.
For TOI-560, the current system parameters indicate that planet--planet secular coupling dominates over the torque from the stellar quadrupole moment. Adopting a stellar quadrupole coefficient of $J_2 \simeq 4.0\times10^{-6}$, estimated assuming a stellar Love number of $k_2=0.28$ following Equation~24 of \citet{Spalding} for a fully convective star, the nodal precession rates induced by planet--planet secular interactions, parameterized by the coupling coefficients $B_{bc}$ and $B_{cb}$ following \citet{Spalding}, are of order $10^{-3}\,{\rm yr}^{-1}$. Using the planet masses reported by \citet{Barragan2022_TOI560}, we obtain $|B_{bc}|=2.05\times10^{-3}\,{\rm yr}^{-1}$ and $|B_{cb}|=1.50\times10^{-3}\,{\rm yr}^{-1}$. In contrast, the nodal precession rates induced by the stellar quadrupole, $\nu_{J2}$, are only $\nu_{J2,b}=5.72\times10^{-6}\,{\rm yr}^{-1}$ and $\nu_{J2,c}=4.60\times10^{-7}\,{\rm yr}^{-1}$ for planets b and c, respectively. These results suggest that the two planets in TOI-560 remain strongly secularly coupled and approximately coplanar, undergoing slow nodal precession as a group rather than differential precession driven by the stellar quadrupole. 
In contrast, TOI-5082 is currently known as a single-transiting system, preventing an equivalent secular-coupling analysis, although the presence of additional non-transiting companions cannot be ruled out.

\begin{figure*}[htbp]
    \centering
    \includegraphics[width=\textwidth]{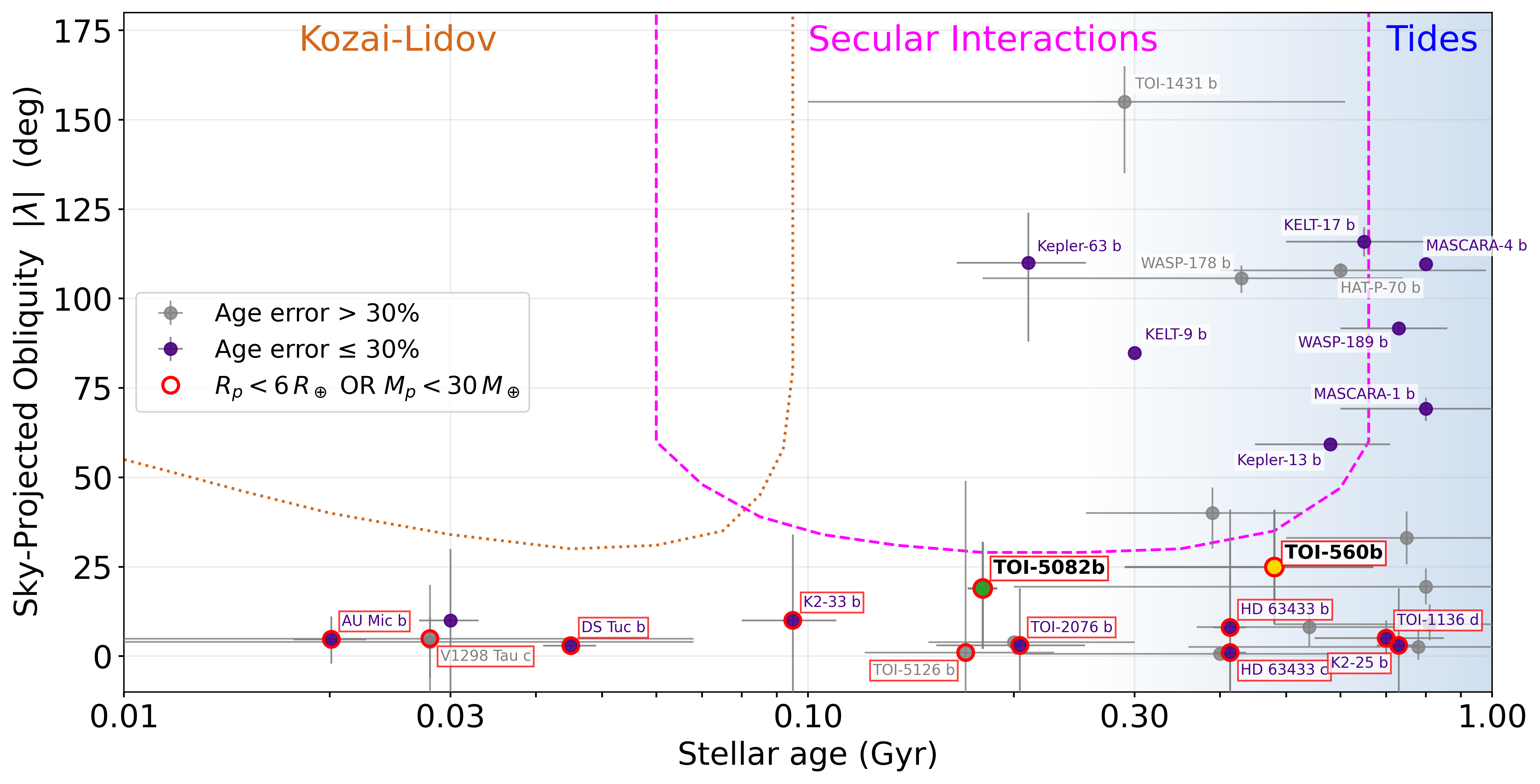}
    \caption{Sky-projected obliquity ($|\lambda|$) as a function of stellar age for systems younger than 1\,Gyr. Systems with well-constrained ages (fractional uncertainty $\leq 30\%$) are shown in purple, while those with poorly constrained ages ($>30\%$) are shown in gray. Planets with radii $R_p < 6\,R_\oplus$ or masses $M_p < 30\,M_\oplus$ are highlighted with red-edged markers and labeled individually. The Kozai--Lidov and secular-interaction contours are adapted from \citet{Zhou2020}, where Kozai interactions can occur over $10^4$--$10^8$\,yr and secular interactions may take over hundreds of Myr. The blue-shaded region highlights the regime in which tidal evolution is expected to become important. Giant planets with strong misalignment ($|\lambda| > 50^\circ$) are also labeled. Our targets, TOI-560\,b (yellow) and TOI-5082\,b (green), are highlighted for comparison.}
    \label{fig:lambda_age}
\end{figure*}

As shown in Figure~\ref{fig:lambda_age}, both TOI-560\,b and TOI-5082\,b are young enough and small enough that tidal realignment is inefficient. Following the equilibrium-tide formalism of \citet{Lai2012}, as summarized by \citet{LiWinn2016}, we estimate angular-momentum-transfer timescales of $4.0^{+5.0}_{-1.8}\times10^{8}$\,Gyr and $2.0^{+7.9}_{-1.3}\times10^{8}$\,Gyr for TOI-560\,b and TOI-5082\,b, respectively, much longer than both the ages of the systems and the age of the Universe. The corresponding ratios of the planetary orbital angular momentum to the stellar spin angular momentum are $L_{\rm orb}/J_\star = 0.52^{+0.17}_{-0.17}$ and $0.12^{+0.07}_{-0.06}$, respectively. Although the orbital angular momentum is not negligible compared to the stellar spin angular momentum, the tidal torque is extremely weak for these small planets at their present orbital separations, so tides are unable to erase any primordial misalignment due to tilting of the protoplanetary disk \citep{Batygin2012,Lai2014}.
Our measured low obliquities therefore do not favor strong primordial misalignment. On the other hand,  TOI-560 and TOI-5082 are also old enough that secular perturbations could have excited non-zero obliquities over their 100s of Myr lifetime \citep[e.g.][]{WuLithwick2011}, yet no significant spin--orbit misalignment is observed. More broadly, the population shown in Figure~\ref{fig:lambda_age} reveals an emerging age-dependent structure. At the youngest ages, before secular interactions are expected to become important, nearly all existing spin-orbit measurements favor low obliquities, consistent with primordial alignment. However, misaligned planets begin to appear at ages of $\sim$300 Myr, especially among single hot Jupiters such as Kepler-63 b \citep{Sanchis_Kepler63} and KELT-9 \citep{Gaudi,Ahlers}. One may argue that the formation of these misaligned, single, hot Jupiters may take hundreds of Myr, which would favor high-eccentricity migration produced by Kozai-Lidov or secular chaos \citep{Fabrycky2007,Wu2007,Naoz2011,WuLithwick2011,Petrovich2015,DawsonJohnson2018}. We caution that a sample of hot Jupiters with precisely measured young ages and stellar obliquities is still missing. One important anchoring point is the 58$\pm$5-Myr-old \citep{Distler} KELT-20 Ab which is on an aligned orbit \citep[$\lambda=3.4\pm2.1^{\circ}$][]{Lund}. We caution that some young sub-Neptunes \citep[e.g. V1298 Tau planets, and the HIP-67522 planets][]{Livingston,Thao} may masquerade as hot Jupiters, since radius inflation can make their radii appear Jupiter-like. Nonetheless, more careful mass measurements revealed that they are really inflated sub-Neptunes \citep{Livingston,Thao,Blunt,Barat}.

Neptune-sized or sub-Neptune planets generally favor aligned configurations \citep{Albrecht2022}, indicating that HEM likely applies only to a limited subset of these systems \citep{Zhou2020,DawsonJohnson2018}. The few well-known misaligned smaller planets, such as GJ~436\,b \citep{GJ436b}, GJ~3470\,b \citep{GJ3470b}, WASP-107\,b \citep{WASP-107b}, HAT-P-11\,b \citep{HAT-P-11b}, TOI-2374\,b \citep{TOI-2374b}, and TOI-1710\,b \citep{TOI-1710b_1,TOI-1710b_2}, likely represent dynamically excited outliers rather than the dominant evolutionary pathway. These systems provide concrete examples of dynamically excited architectures: GJ~436\,b is on a polar orbit with a significant eccentricity, consistent with delayed Kozai migration; GJ~3470\,b is likewise on a polar orbit with mild eccentricity and may be influenced by an outer companion \citep{GJ3470b}; WASP-107\,b occupies a near-polar or retrograde orbit in a system with a known distant companion \citep{WASP-107b,Piaulet2021}; HAT-P-11\,b exhibits a high stellar obliquity that has been interpreted as evidence for few-body dynamical interactions rather than smooth disk-driven migration \citep{HAT-P-11b_Winn2010, HAT-P-11b, Yee2018}; TOI-2374\,b was recently found to reside on a nearly polar orbit within the emerging ``Neptunian ridge'' population \citep{TOI-2374b}; and TOI-1710\,A\,b exhibits a strongly retrograde orbit despite being a tidally detached warm Neptune, suggesting that dynamically excited migration may operate even at longer orbital periods \citep{TOI-1710b_1,TOI-1710b_2}. These planets belong to the emerging regime of the ``hot Neptune ridge'' \citep{Castro-Gonzalez2024}, in which a subset of Neptune-sized planets undergo high-eccentricity migration processes analogous to those of hot Jupiters \citep{Yu_Dai,Castro2026,TOI-2374b,Wang2026WWS}. Similar to hot Jupiters, hot Neptunes are also preferentially found around metal-rich host stars \citep{Dong2018,Dai2021,Vissapragada2025}, further supporting a possible connection between the formation pathways of hot Jupiters and the dynamically excited hot-Neptune population. Taken together, the current misaligned planet sample suggests that secular chaos and Kozai--Lidov mechanisms may produce single, misaligned hot Jupiters or hot Neptunes. However, Kepler-like systems with multiple Neptune-sized or smaller planets appear to be largely aligned \citep{Handley}. The only exception is the K2-290 system where multiple transiting Neptune-sized planets are misaligned \citep{Hjorth}. It is likely that the conditions for strong dynamical excitation are not met in most Kepler-like systems \citep{Zhu}. If such conditions are met, the subsequent dynamical evolution produces a singly transiting, misaligned planet, while the other planets are likely driven out of the transiting plane \citep[`Kepler-Dichotomy'][]{Spalding}.

\section{Summary and Conclusions}
\label{sec:summary}

We presented spin--orbit angle measurements of two Neptune-size planets, TOI-560\,b and TOI-5082\,b, using \textit{TESS} photometry together with time-resolved spectroscopy from the Keck Planet Finder. These systems probe an important age range for planetary evolution, corresponding to the onset of dynamical processes such as secular chaos \citep{WuLithwick2011}.

Using a joint fit to the \textit{TESS} transit light curves and the KPF Rossiter--McLaughlin observations with a modified version of {\tt allesfitter}, we measured the sky-projected stellar obliquities of both systems. For TOI-560\,b, we find a projected obliquity of $\lambda_b = -25 \pm 16^\circ$, while for TOI-5082\,b we obtain $\lambda_b = 19^{+17}_{-13}{}^\circ$. These values indicate that both planets are consistent with low projected obliquities. By combining these RM constraints with stellar rotation periods from photometric modulation analyses and independent spectroscopic estimates of $\vsini$, we further inferred the true three-dimensional obliquities. We obtained 95\% upper limits on the true three-dimensional obliquities of $\psi < 69.7^\circ$ for TOI-560\,b and $\psi < 50.5^\circ$ for TOI-5082\,b. Both systems are consistent with low-to-moderate true obliquities and show no evidence for strong misalignment, although non-zero true obliquities cannot be ruled out.

Both systems lie in the regime where secular interactions could begin to operate (hundreds of Myr), yet neither shows evidence for significant spin--orbit misalignment. When compared with planets in systems younger than $\sim 1$\,Gyr, Neptune-size planets—including TOI-560\,b and TOI-5082\,b—consistently exhibit low obliquities across this age range. In contrast, large obliquities at similar or older ages are predominantly associated with more massive planets or systems more consistent with high-eccentricity migration (single hot Jupiters and single hot Neptunes). These results indicate that substantial spin--orbit misalignment is not a common early outcome for Neptune-size planets. Instead, large obliquities in this regime appear to require additional dynamical conditions, such as long-term perturbations or the presence of external companions. 

\begin{acknowledgments}
We thank Soumyadeep Bhattacharjee for his assistance with the KPF observations of TOI-5082 and for his contributions to the acquisition of the data analyzed in this study.

This work made use of the Exoplanet Follow-up Observing Program for TESS (ExoFOP-TESS), available at \dataset[10.26134/ExoFOP3]{https://doi.org/10.26134/ExoFOP3}.
\end{acknowledgments}

\bibliography{citation}{}

@ARTICLE{McLaughlin1924,
       author = {{McLaughlin}, D.~B.},
        title = "{Some results of a spectrographic study of the Algol system.}",
      journal = {\apj},
         year = 1924,
        month = jul,
       volume = {60},
        pages = {22-31},
          doi = {10.1086/142826},
       adsurl = {https://ui.adsabs.harvard.edu/abs/1924ApJ....60...22M}
}

@ARTICLE{Wang2026WWS,
       author = {{Wang}, Xian-Yu and {Wang}, Songhu},
        title = "{Warm Sub-Saturns Orbiting Single Stars Are Spin-Orbit Aligned}",
      journal = {arXiv e-prints},
         year = 2026,
        month = jul,
          eid = {arXiv:2607.29558},
        pages = {arXiv:2607.29558},
archivePrefix = {arXiv},
       eprint = {2607.29558},
 primaryClass = {astro-ph.EP},
       adsurl = {https://ui.adsabs.harvard.edu/abs/2026arXiv260729558W}
}

@ARTICLE{Bourrier2025ATREIDES,
       author = {{Bourrier}, V. and {Steiner}, M. and {Castro-Gonz{\'a}lez}, A. and {Armstrong}, D.~J. and {Attia}, M. and {Gill}, S. and {Timmermans}, M. and {Fernandez}, J. and {Hawthorn}, F. and {Triaud}, A.~H.~M.~J. and {Murgas}, F. and {Palle}, E. and {Chakraborty}, H. and {Poppenhaeger}, K. and {Lendl}, M. and {Anderson}, D.~R. and {Bryant}, E.~M. and {Friden}, E. and {Seidel}, J.~V. and {Zapatero Osorio}, M.~R. and {Eeles-Nolle}, F. and {Lafarga}, M. and {Lockley}, I.~S. and {Serrano Bell}, J. and {Allart}, R. and {Meech}, A. and {Osborn}, A. and {D{\'\i}az}, R.~F. and {Fetzner Keniger}, M.~A. and {Frame}, G. and {Heitzmann}, A. and {Ringham}, A. and {Eggenberger}, P. and {Alibert}, Y. and {Almenara}, J.~M. and {Leleu}, A. and {Sousa}, S.~G. and {Mercier}, S.~J. and {Adibekyan}, V. and {Battley}, M.~P. and {Delgado Mena}, E. and {Dethier}, W. and {Egger}, J.~A. and {Barkaoui}, K. and {Bayliss}, D. and {Burdanov}, A.~Y. and {Ducrot}, E. and {Ghachoui}, M. and {Gillon}, M. and {G{\'o}mez Maqueo Chew}, Y. and {Jehin}, E. and {Pedersen}, P.~P. and {Pozuelos}, F.~J. and {Wheatley}, P.~J. and {Z{\'u}niga-Fern{\'a}ndez}, S. and {Carteret}, Y. and {Cegla}, H.~M. and {Correia}, A.~C.~M. and {Davis}, Y.~T. and {Doyle}, L. and {Ehrenreich}, D. and {Hara}, N.~C. and {Lavie}, B. and {Lillo-Box}, J. and {Lovis}, C. and {Petit}, A.~C. and {Santos}, N.~C. and {Scott}, M.~G. and {Venturini}, J. and {Ahrer}, E.-M. and {Aigrain}, S. and {Barros}, S.~C.~C. and {Gillen}, E. and {Luo}, X. and {Mordasini}, C. and {Al Moulla}, K. and {Pepe}, F. and {Pietrow}, A.~G.~M.},
        title = "{ATREIDES: I. Embarking on a trek across the exo-Neptunian landscape with the TOI-421 system}",
      journal = {\aap},
         year = 2025,
        month = sep,
       volume = {701},
          eid = {A190},
        pages = {A190},
          doi = {10.1051/0004-6361/202554856},
archivePrefix = {arXiv},
       eprint = {2509.15746},
 primaryClass = {astro-ph.EP},
       adsurl = {https://ui.adsabs.harvard.edu/abs/2025A&A...701A.190B}
}

@ARTICLE{Albrecht2022,
       author = {{Albrecht}, Simon H. and {Dawson}, Rebekah I. and {Winn}, Joshua N.},
        title = "{Stellar Obliquities in Exoplanetary Systems}",
      journal = {\pasp},
         year = 2022,
        month = aug,
       volume = {134},
       number = {1038},
          eid = {082001},
        pages = {082001},
          doi = {10.1088/1538-3873/ac6c09},
archivePrefix = {arXiv},
       eprint = {2203.05460},
 primaryClass = {astro-ph.EP},
       adsurl = {https://ui.adsabs.harvard.edu/abs/2022PASP..134h2001A}
}

@ARTICLE{Owen,
       author = {{Owen}, James E.},
        title = "{Atmospheric Escape and the Evolution of Close-In Exoplanets}",
      journal = {Annual Review of Earth and Planetary Sciences},
         year = 2019,
        month = may,
       volume = {47},
        pages = {67-90},
          doi = {10.1146/annurev-earth-053018-060246},
archivePrefix = {arXiv},
       eprint = {1807.07609},
 primaryClass = {astro-ph.EP},
       adsurl = {https://ui.adsabs.harvard.edu/abs/2019AREPS..47...67O}
}

@ARTICLE{Castro2026,
       author = {{Castro-Gonz{\'a}lez}, A. and {Bourrier}, V. and {Ehrenreich}, D. and {Armstrong}, D.~J. and {Correia}, A.~C.~M. and {Lendl}, M.},
        title = "{The Neptunian ridge as a natural outcome of high-eccentricity tidal migration}",
      journal = {arXiv e-prints},
         year = 2026,
        month = apr,
          eid = {arXiv:2604.16300},
        pages = {arXiv:2604.16300},
          doi = {10.48550/arXiv.2604.16300},
archivePrefix = {arXiv},
       eprint = {2604.16300},
 primaryClass = {astro-ph.EP},
       adsurl = {https://ui.adsabs.harvard.edu/abs/2026arXiv260416300C}
}

@ARTICLE{Distler,
       author = {{Distler}, Adam and {Soares-Furtado}, Melinda and {Mann}, Andrew W. and {Kraus}, Adam L. and {Gagn{\'e}}, Jonathan and {Becker}, Juliette and {Narayan}, Ritvik Sai and {Clark}, Max and {Vanderburg}, Andrew and {Rodriguez}, Joseph E. and {Rogers}, Laura K. and {Kerr}, Ronan},
        title = "{TESS Hunt for Young and Maturing Exoplanets (THYME). XIV. A Comoving-based Age Constraint for KELT-20}",
      journal = {\aj},
         year = 2026,
        month = apr,
       volume = {171},
       number = {4},
          eid = {248},
        pages = {248},
          doi = {10.3847/1538-3881/ae4c48},
archivePrefix = {arXiv},
       eprint = {2603.01313},
 primaryClass = {astro-ph.EP},
       adsurl = {https://ui.adsabs.harvard.edu/abs/2026AJ....171..248D}
}

@ARTICLE{Zhu,
       author = {{Zhu}, Wei and {Dong}, Subo},
        title = "{Exoplanet Statistics and Theoretical Implications}",
      journal = {\araa},
         year = 2021,
        month = sep,
       volume = {59},
        pages = {291-336},
          doi = {10.1146/annurev-astro-112420-020055},
archivePrefix = {arXiv},
       eprint = {2103.02127},
 primaryClass = {astro-ph.EP},
       adsurl = {https://ui.adsabs.harvard.edu/abs/2021ARA&A..59..291Z}
}

@ARTICLE{Lund,
       author = {{Lund}, Michael B. and {Rodriguez}, Joseph E. and {Zhou}, George and {Gaudi}, B. Scott and {Stassun}, Keivan G. and {Johnson}, Marshall C. and {Bieryla}, Allyson and {Oelkers}, Ryan J. and {Stevens}, Daniel J. and {Collins}, Karen A. and {Penev}, Kaloyan and {Quinn}, Samuel N. and {Latham}, David W. and {Villanueva}, Jr., Steven and {Eastman}, Jason D. and {Kielkopf}, John F. and {Oberst}, Thomas E. and {Jensen}, Eric L.~N. and {Cohen}, David H. and {Joner}, Michael D. and {Stephens}, Denise C. and {Relles}, Howard and {Corfini}, Giorgio and {Gregorio}, Joao and {Zambelli}, Roberto and {Esquerdo}, Gilbert A. and {Calkins}, Michael L. and {Berlind}, Perry and {Ciardi}, David R. and {Dressing}, Courtney and {Patel}, Rahul and {Gagnon}, Patrick and {Gonzales}, Erica and {Beatty}, Thomas G. and {Siverd}, Robert J. and {Labadie-Bartz}, Jonathan and {Kuhn}, Rudolf B. and {Col{\'o}n}, Knicole D. and {James}, David and {Pepper}, Joshua and {Fulton}, Benjamin J. and {McLeod}, Kim K. and {Stockdale}, Christopher and {Calchi Novati}, Sebastiano and {DePoy}, D.~L. and {Gould}, Andrew and {Marshall}, Jennifer L. and {Trueblood}, Mark and {Trueblood}, Patricia and {Johnson}, John A. and {Wright}, Jason and {McCrady}, Nate and {Wittenmyer}, Robert A. and {Johnson}, Samson A. and {Sergi}, Anthony and {Wilson}, Maurice and {Sliski}, David H.},
        title = "{KELT-20b: A Giant Planet with a Period of P {\ensuremath{\sim}} 3.5 days Transiting the V {\ensuremath{\sim}} 7.6 Early A Star HD 185603}",
      journal = {\aj},
         year = 2017,
        month = nov,
       volume = {154},
       number = {5},
          eid = {194},
        pages = {194},
          doi = {10.3847/1538-3881/aa8f95},
archivePrefix = {arXiv},
       eprint = {1707.01518},
 primaryClass = {astro-ph.EP},
       adsurl = {https://ui.adsabs.harvard.edu/abs/2017AJ....154..194L}
}

@ARTICLE{Izidoro2017,
       author = {{Izidoro}, Andre and {Ogihara}, Masahiro and {Raymond}, Sean N. and {Morbidelli}, Alessandro and {Pierens}, Arnaud and {Bitsch}, Bertram and {Cossou}, Christophe and {Hersant}, Franck},
        title = "{Breaking the chains: hot super-Earth systems from migration and disruption of compact resonant chains}",
      journal = {\mnras},
         year = 2017,
        month = sep,
       volume = {470},
       number = {2},
        pages = {1750-1770},
          doi = {10.1093/mnras/stx1232},
archivePrefix = {arXiv},
       eprint = {1703.03634},
 primaryClass = {astro-ph.EP},
       adsurl = {https://ui.adsabs.harvard.edu/abs/2017MNRAS.470.1750I}
}

@INPROCEEDINGS{Schmidt,
       author = {{Schmidt}, Stephen and {Schlaufman}, Kevin and {Hamer}, Jacob},
        title = "{Resonant and Ultra-short-period Planet Systems are at Opposite Extremes of the Exoplanet Age Distribution}",
    booktitle = {American Astronomical Society Meeting Abstracts \#243},
         year = 2024,
       series = {American Astronomical Society Meeting Abstracts},
       volume = {243},
        month = feb,
          eid = {462.02},
        pages = {462.02},
       adsurl = {https://ui.adsabs.harvard.edu/abs/2024AAS...24346202S}
}

@ARTICLE{Dai2024,
       author = {{Dai}, Fei and {Goldberg}, Max and {Batygin}, Konstantin and {van Saders}, Jennifer and {Chiang}, Eugene and {Choksi}, Nick and {Li}, Rixin and {Petigura}, Erik A. and {Gilbert}, Gregory J. and {Millholland}, Sarah C. and {Dai}, Yuan-Zhe and {Bouma}, Luke and {Weiss}, Lauren M. and {Winn}, Joshua N.},
        title = "{The Prevalence of Resonance Among Young, Close-in Planets}",
      journal = {\aj},
         year = 2024,
        month = dec,
       volume = {168},
       number = {6},
          eid = {239},
        pages = {239},
          doi = {10.3847/1538-3881/ad83a6},
archivePrefix = {arXiv},
       eprint = {2406.06885},
 primaryClass = {astro-ph.EP},
       adsurl = {https://ui.adsabs.harvard.edu/abs/2024AJ....168..239D}
}

@ARTICLE{Handley,
       author = {{Handley}, Luke B. and {Howard}, Andrew W. and {Dai}, Fei and {Rubenzahl}, Ryan A. and {Giacalone}, Steven and {Isaacson}, Howard and {Ong}, J.~M. Joel and {Carmichael}, Theron W. and {Li}, Yaguang and {Lubin}, Jack and {Premnath}, Pranav H. and {Rogers}, Claire J. and {Nagarajan}, Pranav and {Gilbert}, Gregory J. and {Fulton}, Benjamin and {Gibson}, Steven R. and {Roy}, Arpita and {Edelstein}, Jerry and {Smith}, Christopher},
        title = "{The KPF-SLOPE Survey - Small, Compact Multi-Planet Systems Appear Spin-Orbit Aligned}",
      journal = {arXiv e-prints},
         year = 2026,
        month = mar,
          eid = {arXiv:2603.23713},
        pages = {arXiv:2603.23713},
          doi = {10.48550/arXiv.2603.23713},
archivePrefix = {arXiv},
       eprint = {2603.23713},
 primaryClass = {astro-ph.EP},
       adsurl = {https://ui.adsabs.harvard.edu/abs/2026arXiv260323713H}
}

@ARTICLE{Spalding,
       author = {{Spalding}, Christopher and {Batygin}, Konstantin},
        title = "{Spin-Orbit Misalignment as a Driver of the Kepler Dichotomy}",
      journal = {\apj},
         year = 2016,
        month = oct,
       volume = {830},
       number = {1},
          eid = {5},
        pages = {5},
          doi = {10.3847/0004-637X/830/1/5},
archivePrefix = {arXiv},
       eprint = {1607.03999},
 primaryClass = {astro-ph.EP},
       adsurl = {https://ui.adsabs.harvard.edu/abs/2016ApJ...830....5S}
}

@ARTICLE{Teng,
       author = {{Teng}, Huan-Yu and {Dai}, Fei and {Howard}, Andrew W. and {Halverson}, Samuel and {Isaacson}, Howard and {Kokubo}, Eiichiro and {Rubenzahl}, Ryan A. and {Fulton}, Benjamin and {Householder}, Aaron and {Lubin}, Jack and {Giacalone}, Steven and {Handley}, Luke and {Van Zandt}, Judah and {Petigura}, Erik A. and {Ong}, J.~M. Joel and {Premnath}, Pranav and {Yu}, Haochuan and {Gibson}, Steven R. and {Rider}, Kodi and {Roy}, Arpita and {Baker}, Ashley and {Edelstein}, Jerry and {Smith}, Chris and {Walawender}, Josh and {Lee}, Byeong-Cheol and {Liu}, Yu-Juan and {Winn}, Joshua N.},
        title = "{Stellar Obliquity of the Ultra-short-period Planet System HD 93963}",
      journal = {\aj},
         year = 2025,
        month = jul,
       volume = {170},
       number = {1},
          eid = {51},
        pages = {51},
          doi = {10.3847/1538-3881/addab9},
archivePrefix = {arXiv},
       eprint = {2505.10804},
 primaryClass = {astro-ph.EP},
       adsurl = {https://ui.adsabs.harvard.edu/abs/2025AJ....170...51T}
}

@ARTICLE{Ahlers,
       author = {{Ahlers}, John P. and {Johnson}, Marshall C. and {Stassun}, Keivan G. and {Col{\'o}n}, Knicole D. and {Barnes}, Jason W. and {Stevens}, Daniel J. and {Beatty}, Thomas and {Gaudi}, B. Scott and {Collins}, Karen A. and {Rodriguez}, Joseph E. and {Ricker}, George and {Vanderspek}, Roland and {Latham}, David and {Seager}, Sara and {Winn}, Joshua and {Jenkins}, Jon M. and {Caldwell}, Douglas A. and {Goeke}, Robert F. and {Osborn}, Hugh P. and {Paegert}, Martin and {Rowden}, Pam and {Tenenbaum}, Peter},
        title = "{KELT-9 b's Asymmetric TESS Transit Caused by Rapid Stellar Rotation and Spin-Orbit Misalignment}",
      journal = {\aj},
         year = 2020,
        month = jul,
       volume = {160},
       number = {1},
          eid = {4},
        pages = {4},
          doi = {10.3847/1538-3881/ab8fa3},
archivePrefix = {arXiv},
       eprint = {2004.14812},
 primaryClass = {astro-ph.EP},
       adsurl = {https://ui.adsabs.harvard.edu/abs/2020AJ....160....4A}
}

@ARTICLE{Yu_Dai,
       author = {{Yu}, Hang and {Dai}, Fei},
        title = "{Are WASP-107-like Systems Consistent with High-eccentricity Migration?}",
      journal = {\apj},
         year = 2024,
        month = sep,
       volume = {972},
       number = {2},
          eid = {159},
        pages = {159},
          doi = {10.3847/1538-4357/ad5ffb},
archivePrefix = {arXiv},
       eprint = {2406.00187},
 primaryClass = {astro-ph.EP},
       adsurl = {https://ui.adsabs.harvard.edu/abs/2024ApJ...972..159Y}
}

@ARTICLE{Barat,
       author = {{Barat}, Saugata and {D{\'e}sert}, Jean-Michel and {Mukherjee}, Sagnick and {Goyal}, Jayesh M. and {Xue}, Qiao and {Kawashima}, Yui and {Vazan}, Allona and {Misener}, William and {Schlichting}, Hilke E. and {Fortney}, Jonathan J. and {Bean}, Jacob L. and {Avarsekar}, Swaroop and {Henry}, Gregory W. and {Baeyens}, Robin and {Line}, Michael R. and {Livingston}, John H. and {David}, Trevor and {Petigura}, Erik A. and {Sikora}, James T. and {Shivkumar}, Hinna and {Feinstein}, Adina D. and {Oklop{\v{c}}i{\'c}}, Antonija},
        title = "{A Metal-poor Atmosphere with a Hot Interior for a Young Sub-Neptune Progenitor: JWST/NIRSpec Transmission Spectrum of V1298 Tau b}",
      journal = {\aj},
         year = 2025,
        month = sep,
       volume = {170},
       number = {3},
          eid = {165},
        pages = {165},
          doi = {10.3847/1538-3881/adec89},
archivePrefix = {arXiv},
       eprint = {2507.08837},
 primaryClass = {astro-ph.EP},
       adsurl = {https://ui.adsabs.harvard.edu/abs/2025AJ....170..165B}
}

@ARTICLE{Blunt,
       author = {{Blunt}, Sarah and {Carvalho}, Adolfo and {David}, Trevor J. and {Beichman}, Charles and {Zink}, Jon K. and {Gaidos}, Eric and {Behmard}, Aida and {Bouma}, Luke G. and {Cody}, Devin and {Dai}, Fei and {Foreman-Mackey}, Daniel and {Grunblatt}, Sam and {Howard}, Andrew W. and {Kosiarek}, Molly and {Knutson}, Heather A. and {Rubenzahl}, Ryan A. and {Beard}, Corey and {Chontos}, Ashley and {Giacalone}, Steven and {Hirano}, Teruyuki and {Johnson}, Marshall C. and {Lubin}, Jack and {Akana Murphy}, Joseph M. and {Petigura}, Erik A. and {Van Zandt}, Judah and {Weiss}, Lauren},
        title = "{Overfitting Affects the Reliability of Radial Velocity Mass Estimates of the V1298 Tau Planets}",
      journal = {\aj},
         year = 2023,
        month = aug,
       volume = {166},
       number = {2},
          eid = {62},
        pages = {62},
          doi = {10.3847/1538-3881/acde78},
archivePrefix = {arXiv},
       eprint = {2306.08145},
 primaryClass = {astro-ph.EP},
       adsurl = {https://ui.adsabs.harvard.edu/abs/2023AJ....166...62B}
}

@ARTICLE{Hjorth,
       author = {{Hjorth}, Maria and {Albrecht}, Simon and {Hirano}, Teruyuki and {Winn}, Joshua N. and {Dawson}, Rebekah I. and {Zanazzi}, J.~J. and {Knudstrup}, Emil and {Sato}, Bun'ei},
        title = "{A backward-spinning star with two coplanar planets}",
      journal = {Proceedings of the National Academy of Science},
         year = 2021,
        month = feb,
       volume = {118},
       number = {8},
          eid = {e2017418118},
        pages = {e2017418118},
          doi = {10.1073/pnas.2017418118},
archivePrefix = {arXiv},
       eprint = {2102.07677},
 primaryClass = {astro-ph.EP},
       adsurl = {https://ui.adsabs.harvard.edu/abs/2021PNAS..11817418H}
}

@ARTICLE{Thao,
       author = {{Thao}, Pa Chia and {Mann}, Andrew W. and {Feinstein}, Adina D. and {Gao}, Peter and {Thorngren}, Daniel and {Rotman}, Yoav and {Welbanks}, Luis and {Brown}, Alexander and {Duvvuri}, Girish M. and {France}, Kevin and {Longo}, Isabella and {Sandoval}, Angeli and {Schneider}, P. Christian and {Wilson}, David J. and {Youngblood}, Allison and {Vanderburg}, Andrew and {Barber}, Madyson G. and {Wood}, Mackenna L. and {Batalha}, Natasha E. and {Kraus}, Adam L. and {Murray}, Catriona Anne and {Newton}, Elisabeth R. and {Rizzuto}, Aaron and {Tofflemire}, Benjamin M. and {Tsai}, Shang-Min and {Bean}, Jacob L. and {Berta-Thompson}, Zachory K. and {Evans-Soma}, Thomas M. and {Froning}, Cynthia S. and {Kempton}, Eliza M.-R. and {Miguel}, Yamila and {Pineda}, J. Sebastian},
        title = "{The Featherweight Giant: Unraveling the Atmosphere of a 17 Myr Planet with JWST}",
      journal = {\aj},
         year = 2024,
        month = dec,
       volume = {168},
       number = {6},
          eid = {297},
        pages = {297},
          doi = {10.3847/1538-3881/ad81d7},
archivePrefix = {arXiv},
       eprint = {2409.16355},
 primaryClass = {astro-ph.EP},
       adsurl = {https://ui.adsabs.harvard.edu/abs/2024AJ....168..297T}
}

@ARTICLE{Livingston,
       author = {{Livingston}, John H. and {Petigura}, Erik A. and {David}, Trevor J. and {Masuda}, Kento and {Owen}, James and {Nesvorn{\'y}}, David and {Batygin}, Konstantin and {de Leon}, Jerome and {Mori}, Mayuko and {Ikuta}, Kai and {Fukui}, Akihiko and {Watanabe}, Noriharu and {Orell Miquel}, Jaume and {Murgas}, Felipe and {Parviainen}, Hannu and {Korth}, Judith and {Libotte}, Florence and {Abreu Garc{\'\i}a}, N{\'e}stor and {Gallardo}, Pedro Pablo Meni and {Narita}, Norio and {Pall{\'e}}, Enric and {Tamura}, Motohide and {Yonehara}, Atsunori and {Ridden-Harper}, Andrew and {Bieryla}, Allyson and {Trani}, Alessandro A. and {Mamajek}, Eric E. and {Ciardi}, David R. and {Gorjian}, Varoujan and {Hillenbrand}, Lynne A. and {Rebull}, Luisa M. and {Newton}, Elisabeth R. and {Mann}, Andrew W. and {Vanderburg}, Andrew and {Stef{\'a}nsson}, Gu{\dh}mundur and {Mahadevan}, Suvrath and {Ca{\~n}as}, Caleb and {Ninan}, Joe and {Higuera}, Jesus and {Todorov}, Kamen and {D{\'e}sert}, Jean-Michel and {Pino}, Lorenzo},
        title = "{A young progenitor for the most common planetary systems in the Galaxy}",
      journal = {\nat},
         year = 2026,
        month = jan,
       volume = {649},
       number = {8096},
        pages = {310-314},
          doi = {10.1038/s41586-025-09840-z},
archivePrefix = {arXiv},
       eprint = {2601.10598},
 primaryClass = {astro-ph.EP},
       adsurl = {https://ui.adsabs.harvard.edu/abs/2026Natur.649..310L}
}

@ARTICLE{Gaudi,
       author = {{Gaudi}, B. Scott and {Stassun}, Keivan G. and {Collins}, Karen A. and {Beatty}, Thomas G. and {Zhou}, George and {Latham}, David W. and {Bieryla}, Allyson and {Eastman}, Jason D. and {Siverd}, Robert J. and {Crepp}, Justin R. and {Gonzales}, Erica J. and {Stevens}, Daniel J. and {Buchhave}, Lars A. and {Pepper}, Joshua and {Johnson}, Marshall C. and {Colon}, Knicole D. and {Jensen}, Eric L.~N. and {Rodriguez}, Joseph E. and {Bozza}, Valerio and {Novati}, Sebastiano Calchi and {D'Ago}, Giuseppe and {Dumont}, Mary T. and {Ellis}, Tyler and {Gaillard}, Clement and {Jang-Condell}, Hannah and {Kasper}, David H. and {Fukui}, Akihiko and {Gregorio}, Joao and {Ito}, Ayaka and {Kielkopf}, John F. and {Manner}, Mark and {Matt}, Kyle and {Narita}, Norio and {Oberst}, Thomas E. and {Reed}, Phillip A. and {Scarpetta}, Gaetano and {Stephens}, Denice C. and {Yeigh}, Rex R. and {Zambelli}, Roberto and {Fulton}, B.~J. and {Howard}, Andrew W. and {James}, David J. and {Penny}, Matthew and {Bayliss}, Daniel and {Curtis}, Ivan A. and {Depoy}, D.~L. and {Esquerdo}, Gilbert A. and {Gould}, Andrew and {Joner}, Michael D. and {Kuhn}, Rudolf B. and {Labadie-Bartz}, Jonathan and {Lund}, Michael B. and {Marshall}, Jennifer L. and {McLeod}, Kim K. and {Pogge}, Richard W. and {Relles}, Howard and {Stockdale}, Christopher and {Tan}, T.~G. and {Trueblood}, Mark and {Trueblood}, Patricia},
        title = "{A giant planet undergoing extreme-ultraviolet irradiation by its hot massive-star host}",
      journal = {\nat},
         year = 2017,
        month = jun,
       volume = {546},
       number = {7659},
        pages = {514-518},
          doi = {10.1038/nature22392},
archivePrefix = {arXiv},
       eprint = {1706.06723},
 primaryClass = {astro-ph.EP},
       adsurl = {https://ui.adsabs.harvard.edu/abs/2017Natur.546..514G}
}

@ARTICLE{Sanchis_Kepler63,
       author = {{Sanchis-Ojeda}, Roberto and {Winn}, Joshua N. and {Marcy}, Geoffrey W. and {Howard}, Andrew W. and {Isaacson}, Howard and {Johnson}, John Asher and {Torres}, Guillermo and {Albrecht}, Simon and {Campante}, Tiago L. and {Chaplin}, William J. and {Davies}, Guy R. and {Lund}, Mikkel N. and {Carter}, Joshua A. and {Dawson}, Rebekah I. and {Buchhave}, Lars A. and {Everett}, Mark E. and {Fischer}, Debra A. and {Geary}, John C. and {Gilliland}, Ronald L. and {Horch}, Elliott P. and {Howell}, Steve B. and {Latham}, David W.},
        title = "{Kepler-63b: A Giant Planet in a Polar Orbit around a Young Sun-like Star}",
      journal = {\apj},
         year = 2013,
        month = sep,
       volume = {775},
       number = {1},
          eid = {54},
        pages = {54},
          doi = {10.1088/0004-637X/775/1/54},
archivePrefix = {arXiv},
       eprint = {1307.8128},
 primaryClass = {astro-ph.EP},
       adsurl = {https://ui.adsabs.harvard.edu/abs/2013ApJ...775...54S}
}

@ARTICLE{Zhang2025,
       author = {{Zhang}, Elina Y. and {Teng}, Huan-Yu and {Dai}, Fei and {Howard}, Andrew W. and {Halverson}, Samuel P. and {Isaacson}, Howard and {Rubenzahl}, Ryan A. and {Wang}, Xian-Yu and {Wang}, Songhu and {Fulton}, Benjamin J. and {Nielsen}, Louise D. and {Lubin}, Jack and {Giacalone}, Steven and {Handley}, Luke B. and {Petigura}, Erik A. and {Turtelboom}, Emma V. and {Polanski}, Alex S. and {Gibson}, Steve R. and {Rider}, Kodi and {Roy}, Arpita and {Baker}, Ashley and {Edelstein}, Jerry and {Smith}, Christopher L. and {Walawender}, Josh and {Winn}, Joshua N.},
        title = "{TOI-880 is an Aligned, Coplanar, Multiplanet System}",
      journal = {\aj},
         year = 2025,
        month = sep,
       volume = {170},
       number = {3},
          eid = {175},
        pages = {175},
          doi = {10.3847/1538-3881/adf339},
archivePrefix = {arXiv},
       eprint = {2507.16194},
 primaryClass = {astro-ph.EP},
       adsurl = {https://ui.adsabs.harvard.edu/abs/2025AJ....170..175Z}
}

@ARTICLE{Winn,
       author = {{Winn}, Joshua N. and {Fabrycky}, Daniel C.},
        title = "{The Occurrence and Architecture of Exoplanetary Systems}",
      journal = {\araa},
         year = 2015,
        month = aug,
       volume = {53},
        pages = {409-447},
          doi = {10.1146/annurev-astro-082214-122246},
archivePrefix = {arXiv},
       eprint = {1410.4199},
 primaryClass = {astro-ph.EP},
       adsurl = {https://ui.adsabs.harvard.edu/abs/2015ARA&A..53..409W}
}

@ARTICLE{Petigura,
       author = {{Petigura}, Erik A. and {Howard}, Andrew W. and {Marcy}, Geoffrey W.},
        title = "{Prevalence of Earth-size planets orbiting Sun-like stars}",
      journal = {Proceedings of the National Academy of Science},
         year = 2013,
        month = nov,
       volume = {110},
       number = {48},
        pages = {19273-19278},
          doi = {10.1073/pnas.1319909110},
archivePrefix = {arXiv},
       eprint = {1311.6806},
 primaryClass = {astro-ph.EP},
       adsurl = {https://ui.adsabs.harvard.edu/abs/2013PNAS..11019273P}
}

@ARTICLE{Fressin,
       author = {{Fressin}, Fran{\c{c}}ois and {Torres}, Guillermo and {Charbonneau}, David and {Bryson}, Stephen T. and {Christiansen}, Jessie and {Dressing}, Courtney D. and {Jenkins}, Jon M. and {Walkowicz}, Lucianne M. and {Batalha}, Natalie M.},
        title = "{The False Positive Rate of Kepler and the Occurrence of Planets}",
      journal = {\apj},
         year = 2013,
        month = apr,
       volume = {766},
       number = {2},
          eid = {81},
        pages = {81},
          doi = {10.1088/0004-637X/766/2/81},
archivePrefix = {arXiv},
       eprint = {1301.0842},
 primaryClass = {astro-ph.EP},
       adsurl = {https://ui.adsabs.harvard.edu/abs/2013ApJ...766...81F}
}

@ARTICLE{Rossiter1924,
       author = {{Rossiter}, R.~A.},
        title = "{On the detection of an effect of rotation during eclipse in the velocity of the brigher component of beta Lyrae, and on the constancy of velocity of this system.}",
      journal = {\apj},
         year = 1924,
        month = jul,
       volume = {60},
        pages = {15-21},
          doi = {10.1086/142825},
       adsurl = {https://ui.adsabs.harvard.edu/abs/1924ApJ....60...15R}
}

@misc{isoclassify_code,
       author = {{Huber}, Daniel},
        title = "{isoclassify: v1.2}",
         year = 2017,
        month = may,
          eid = {10.5281/zenodo.573372},
          doi = {10.5281/zenodo.573372},
      version = {v1.2},
    publisher = {Zenodo},
       adsurl = {https://ui.adsabs.harvard.edu/abs/2017zndo....573372H}
}

@ARTICLE{Huber2017,
       author = {{Huber}, Daniel and {Zinn}, Joel and {Bojsen-Hansen}, Mathias and {Pinsonneault}, Marc and {Sahlholdt}, Christian and {Serenelli}, Aldo and {Silva Aguirre}, Victor and {Stassun}, Keivan and {Stello}, Dennis and {Tayar}, Jamie and {Bastien}, Fabienne and {Bedding}, Timothy R. and {Buchhave}, Lars A. and {Chaplin}, William J. and {Davies}, Guy R. and {Garc{\'\i}a}, Rafael A. and {Latham}, David W. and {Mathur}, Savita and {Mosser}, Benoit and {Sharma}, Sanjib},
        title = "{Asteroseismology and Gaia: Testing Scaling Relations Using 2200 Kepler  Stars with TGAS Parallaxes}",
      journal = {\apj},
         year = 2017,
        month = aug,
       volume = {844},
       number = {2},
          eid = {102},
        pages = {102},
          doi = {10.3847/1538-4357/aa75ca},
archivePrefix = {arXiv},
       eprint = {1705.04697},
 primaryClass = {astro-ph.SR},
       adsurl = {https://ui.adsabs.harvard.edu/abs/2017ApJ...844..102H}
}

@ARTICLE{Berger2020,
       author = {{Berger}, Travis A. and {Huber}, Daniel and {van Saders}, Jennifer L. and {Gaidos}, Eric and {Tayar}, Jamie and {Kraus}, Adam L.},
        title = "{The Gaia-Kepler Stellar Properties Catalog. I. Homogeneous Fundamental Properties for 186,301 Kepler Stars}",
      journal = {\aj},
         year = 2020,
        month = jun,
       volume = {159},
       number = {6},
          eid = {280},
        pages = {280},
          doi = {10.3847/1538-3881/159/6/280},
archivePrefix = {arXiv},
       eprint = {2001.07737},
 primaryClass = {astro-ph.SR},
       adsurl = {https://ui.adsabs.harvard.edu/abs/2020AJ....159..280B}
}

@ARTICLE{Berger2023,
       author = {{Berger}, Travis A. and {Schlieder}, Joshua E. and {Huber}, Daniel},
        title = "{The Gaia-Kepler-TESS-Host Stellar Properties Catalog: Uniform Physical Parameters for 7993 Host Stars and 9324 Planets}",
      journal = {arXiv e-prints},
         year = 2023,
        month = jan,
          eid = {arXiv:2301.11338},
        pages = {arXiv:2301.11338},
          doi = {10.48550/arXiv.2301.11338},
archivePrefix = {arXiv},
       eprint = {2301.11338},
 primaryClass = {astro-ph.EP},
       adsurl = {https://ui.adsabs.harvard.edu/abs/2023arXiv230111338B}
}

@ARTICLE{MIST,
       author = {{Choi}, Jieun and {Dotter}, Aaron and {Conroy}, Charlie and {Cantiello}, Matteo and {Paxton}, Bill and {Johnson}, Benjamin D.},
        title = "{Mesa Isochrones and Stellar Tracks (MIST). I. Solar-scaled Models}",
      journal = {\apj},
         year = 2016,
        month = jun,
       volume = {823},
       number = {2},
          eid = {102},
        pages = {102},
          doi = {10.3847/0004-637X/823/2/102},
archivePrefix = {arXiv},
       eprint = {1604.08592},
 primaryClass = {astro-ph.SR},
       adsurl = {https://ui.adsabs.harvard.edu/abs/2016ApJ...823..102C}
}

@ARTICLE{Lindegren2021,
       author = {{Lindegren}, L. and {Bastian}, U. and {Biermann}, M. and {Bombrun}, A. and {de Torres}, A. and {Gerlach}, E. and {Geyer}, R. and {Hern{\'a}ndez}, J. and {Hilger}, T. and {Hobbs}, D. and {Klioner}, S.~A. and {Lammers}, U. and {McMillan}, P.~J. and {Ramos-Lerate}, M. and {Steidelm{\"u}ller}, H. and {Stephenson}, C.~A. and {van Leeuwen}, F.},
        title = "{Gaia Early Data Release 3. Parallax bias versus magnitude, colour, and position}",
      journal = {\aap},
         year = 2021,
        month = may,
       volume = {649},
          eid = {A4},
        pages = {A4},
          doi = {10.1051/0004-6361/202039653},
archivePrefix = {arXiv},
       eprint = {2012.01742},
 primaryClass = {astro-ph.IM},
       adsurl = {https://ui.adsabs.harvard.edu/abs/2021A&A...649A...4L}
}

@ARTICLE{Tayar2022,
       author = {{Tayar}, Jamie and {Claytor}, Zachary R. and {Huber}, Daniel and {van Saders}, Jennifer},
        title = "{A Guide to Realistic Uncertainties on the Fundamental Properties of Solar-type Exoplanet Host Stars}",
      journal = {\apj},
         year = 2022,
        month = mar,
       volume = {927},
       number = {1},
          eid = {31},
        pages = {31},
          doi = {10.3847/1538-4357/ac4bbc},
archivePrefix = {arXiv},
       eprint = {2012.07957},
 primaryClass = {astro-ph.EP},
       adsurl = {https://ui.adsabs.harvard.edu/abs/2022ApJ...927...31T}
}

@ARTICLE{Bovy2016,
       author = {{Bovy}, Jo and {Rix}, Hans-Walter and {Green}, Gregory M. and {Schlafly}, Edward F. and {Finkbeiner}, Douglas P.},
        title = "{On Galactic Density Modeling in the Presence of Dust Extinction}",
      journal = {\apj},
         year = 2016,
        month = feb,
       volume = {818},
       number = {2},
          eid = {130},
        pages = {130},
          doi = {10.3847/0004-637X/818/2/130},
archivePrefix = {arXiv},
       eprint = {1509.06751},
 primaryClass = {astro-ph.GA},
       adsurl = {https://ui.adsabs.harvard.edu/abs/2016ApJ...818..130B}
}

@INPROCEEDINGS{Jenkins2016,
       author = {{Jenkins}, Jon M. and {Twicken}, Joseph D. and {McCauliff}, Sean and {Campbell}, Jennifer and {Sanderfer}, Dwight and {Lung}, David and {Mansouri-Samani}, Masoud and {Girouard}, Forrest and {Tenenbaum}, Peter and {Klaus}, Todd and {Smith}, Jeffrey C. and {Caldwell}, Douglas A. and {Chacon}, A.~D. and {Henze}, Christopher and {Heiges}, Cory and {Latham}, David W. and {Morgan}, Edward and {Swade}, Daryl and {Rinehart}, Stephen and {Vanderspek}, Roland},
        title = "{The TESS science processing operations center}",
    booktitle = {Software and Cyberinfrastructure for Astronomy IV},
         year = 2016,
       editor = {{Chiozzi}, Gianluca and {Guzman}, Juan C.},
       series = {Society of Photo-Optical Instrumentation Engineers (SPIE) Conference Series},
       volume = {9913},
        month = aug,
          eid = {99133E},
        pages = {99133E},
          doi = {10.1117/12.2233418},
       adsurl = {https://ui.adsabs.harvard.edu/abs/2016SPIE.9913E..3EJ}
}

@article{Stumpe2014,
doi = {10.1086/674989},
url = {https://dx.doi.org/10.1086/674989},
year = {2014},
month = {jan},
publisher = {University of Chicago Press},
volume = {126},
number = {935},
pages = {100},
author = {Martin C. Stumpe and Jeffrey C. Smith and Joseph H. Catanzarite and Jeffrey E. Van Cleve and Jon M. Jenkins and Joseph D. Twicken and Forrest R. Girouard},
title = {Multiscale Systematic Error Correction via Wavelet-Based Bandsplitting in Kepler Data},
journal = {Publications of the Astronomical Society of the Pacific},
}

@misc{lightkurve2018,
       author = {{Lightkurve Collaboration} and {Cardoso}, Jos{\'e} Vin{\'\i}cius de Miranda and {Hedges}, Christina and {Gully-Santiago}, Michael and {Saunders}, Nicholas and {Cody}, Ann Marie and {Barclay}, Thomas and {Hall}, Oliver and {Sagear}, Sheila and {Turtelboom}, Emma and {Zhang}, Johnny and {Tzanidakis}, Andy and {Mighell}, Ken and {Coughlin}, Jeff and {Bell}, Keaton and {Berta-Thompson}, Zach and {Williams}, Peter and {Dotson}, Jessie and {Barentsen}, Geert},
        title = "{Lightkurve: Kepler and TESS time series analysis in Python}",
 howpublished = {Astrophysics Source Code Library, record ascl:1812.013},
         year = 2018,
        month = dec,
          eid = {ascl:1812.013},
       adsurl = {https://ui.adsabs.harvard.edu/abs/2018ascl.soft12013L}
}

@article{Smith2012,
doi = {10.1086/667697},
url = {https://dx.doi.org/10.1086/667697},
year = {2012},
month = {sep},
publisher = {University of Chicago Press},
volume = {124},
number = {919},
pages = {1000},
author = {Jeffrey C. Smith and Martin C. Stumpe and Jeffrey E. Van Cleve and Jon M. Jenkins and Thomas S. Barclay and Michael N. Fanelli and Forrest R. Girouard and Jeffery J. Kolodziejczak and Sean D. McCauliff and Robert L. Morris and Joseph D. Twicken},
title = {Kepler Presearch Data Conditioning II - A Bayesian Approach to Systematic Error Correction},
journal = {Publications of the Astronomical Society of the Pacific},
}

@inproceedings{Gibson2024,
author = {Steven R. Gibson and Andrew W. Howard and Kodi Rider and Samuel Halverson and Arpita Roy and Ashley D. Baker and Jerry Edelstein and Christopher Smith and Benjamin J. Fulton and Josh Walawender and Max Brodheim and Matthew Brown and Dwight Chan and Fei Dai and William Deich and Colby Gottschalk and Jason Grillo and David Hale and Grant Hill and Bradford Holden and Aaron Householder and Howard Isaacson and Yuzo Ishikawa and Sharon Jelinsky and Marc Kassis and Stephen Kaye and Russ Laher and Kyle Lanclos and Chien-Hsiu Lee and Scott Lilley and Benjamin McCarney and Timothy N. Miller and Joel Payne and Erik Petigura and Claire Poppett and Michael P. Raffanti and Ryan Rubenzahl and Dale Sandford and Christian Schwab and Abby P. Shaum and Martin M. Sirk and Roger Smith and Jim Thorne and John Valliant and Adam Vandenberg and Shin-Ywan Wang and Edward H. Wishnow and Truman Wold and Sherry Yeh and Steve Baca and Charles Beichman and Bruce Berriman and Thomas Brown and Kelleen Casey and Jason Chin and James Chong and David Cowley and Mark Devenot and Hamza Elwir and Daniel Finstad and Matthew Fraysse and Ean James and Elisha Jhoti and Joe Killian and Obie Levine and Adela Chenyang Li and Eduardo Marin and Steven Milner and Craig Nance and Timothy J. O'Hanlon and Daniel Orr and Roberto Ortiz-Soto and Tom Payne and Jacob Pember and Gert Raskin and Maureen Savage and Andreas Seifahrt and Brett Smith and Rob Storesund and Julian St{\"u}rmer and Nick Suominen and Jerez Tehero and Tod Von Boeckmann and Keith Wages and Marie Weisfeiler and Mavourneen Wilcox and Peter Wizinowich and Anna Wolfenberger},
title = {{System design of the Keck Planet Finder}},
volume = {13096},
booktitle = {Ground-based and Airborne Instrumentation for Astronomy X},
editor = {Julia J. Bryant and Kentaro Motohara and Jo{\"e}l R. D. Vernet},
organization = {International Society for Optics and Photonics},
publisher = {SPIE},
pages = {1309609},
year = {2024},
doi = {10.1117/12.3017841},
URL = {https://doi.org/10.1117/12.3017841}
}

@INPROCEEDINGS{Gibson2020,
       author = {{Gibson}, Steven R. and {Howard}, Andrew W. and {Rider}, Kodi and {Roy}, Arpita and {Edelstein}, Jerry and {Kassis}, Marc and {Grillo}, Jason and {Halverson}, Sam and {Sirk}, Martin M. and {Smith}, Christopher and {Allen}, Steve and {Baker}, Ashley and {Beichman}, Charles and {Berriman}, Bruce and {Brown}, Thomas and {Casey}, Kelleen and {Chin}, Jason and {Coutts}, David and {Cowley}, David and {Deich}, William and {Feger}, Tobias and {Fulton}, Benjamin and {Gers}, Luke and {Gurevich}, Yulia and {Ishikawa}, Yuzo and {James}, Ean and {Jelinsky}, Sharon and {Kaye}, Stephen and {Lanclos}, Kyle and {Li}, Adela and {Lilley}, Scott and {McCarney}, Ben and {Miller}, Tim and {Milner}, Steve and {O'Hanlon}, Timothy J. and {Pember}, Jacob and {Raffanti}, Mike and {Rockosi}, Constance and {Rubenzahl}, Ryan and {Rumph}, Dave and {Sandford}, Dale and {Savage}, Maureen and {Schwab}, Christian and {Seifahrt}, Andreas and {Shaum}, Abby and {Smith}, Roger and {Stuermer}, Julian and {Thorne}, Jim and {Vandenberg}, Adam and {Von Boeckmann}, Tod and {Wang}, Cindy and {Wang}, Qifan and {Weisfeiler}, Marie and {Wilcox}, Mavourneen and {Wishnow}, Edward H. and {Wizinowich}, Peter and {Wold}, Truman and {Wolfenberger}, Anna},
        title = "{Keck Planet Finder: design updates}",
    booktitle = {Ground-based and Airborne Instrumentation for Astronomy VIII},
         year = 2020,
       editor = {{Evans}, Christopher J. and {Bryant}, Julia J. and {Motohara}, Kentaro},
       series = {Society of Photo-Optical Instrumentation Engineers (SPIE) Conference Series},
       volume = {11447},
        month = dec,
          eid = {1144742},
        pages = {1144742},
          doi = {10.1117/12.2561783},
       adsurl = {https://ui.adsabs.harvard.edu/abs/2020SPIE11447E..42G}
}

@INPROCEEDINGS{Gibson2018,
       author = {{Gibson}, Steven R. and {Howard}, Andrew W. and {Roy}, Arpita and {Smith}, Christopher and {Halverson}, Sam and {Edelstein}, Jerry and {Kassis}, Marc and {Wishnow}, Edward H. and {Raffanti}, Michael and {Allen}, Steve and {Chin}, Jason and {Coutts}, David and {Cowley}, David and {Curtis}, Jim and {Deich}, William and {Feger}, Tobias and {Finstad}, Daniel and {Gurevich}, Yulia and {Ishikawa}, Yuzo and {James}, Ean and {Jhoti}, Elisha and {Lanclos}, Kyle and {Lilley}, Scott and {Miller}, Tim and {Milner}, Steve and {Payne}, Tom and {Rider}, Kodi and {Rockosi}, Constance and {Sandford}, Dale and {Schwab}, Christian and {Seifahrt}, Andreas and {Sirk}, Martin M. and {Smith}, Roger and {Stuermer}, Julian and {Weisfeiler}, Marie and {Wilcox}, Mavourneen and {Vandenberg}, Adam and {Wizinowich}, Peter},
        title = "{Keck Planet Finder: preliminary design}",
    booktitle = {Ground-based and Airborne Instrumentation for Astronomy VII},
         year = 2018,
       editor = {{Evans}, Christopher J. and {Simard}, Luc and {Takami}, Hideki},
       series = {Society of Photo-Optical Instrumentation Engineers (SPIE) Conference Series},
       volume = {10702},
        month = jul,
          eid = {107025X},
        pages = {107025X},
          doi = {10.1117/12.2311565},
       adsurl = {https://ui.adsabs.harvard.edu/abs/2018SPIE10702E..5XG}
}

@INPROCEEDINGS{Gibson2016,
       author = {{Gibson}, Steven R. and {Howard}, Andrew W. and {Marcy}, Geoffrey W. and {Edelstein}, Jerry and {Wishnow}, Edward H. and {Poppett}, Claire L.},
        title = "{KPF: Keck Planet Finder}",
    booktitle = {Ground-based and Airborne Instrumentation for Astronomy VI},
         year = 2016,
       editor = {{Evans}, Christopher J. and {Simard}, Luc and {Takami}, Hideki},
       series = {Society of Photo-Optical Instrumentation Engineers (SPIE) Conference Series},
       volume = {9908},
        month = aug,
          eid = {990870},
        pages = {990870},
          doi = {10.1117/12.2233334},
       adsurl = {https://ui.adsabs.harvard.edu/abs/2016SPIE.9908E..70G}
}

@ARTICLE{Rubenzahl2023,
       author = {{Rubenzahl}, Ryan A. and {Halverson}, Samuel and {Walawender}, Josh and {Hill}, Grant M. and {Howard}, Andrew W. and {Brown}, Matthew and {Ida}, Evan and {Tehero}, Jerez and {Fulton}, Benjamin J. and {Gibson}, Steven R. and {Kassis}, Marc and {Smith}, Brett and {Wold}, Truman and {Payne}, Joel},
        title = "{Staring at the Sun with the Keck Planet Finder: An Autonomous Solar Calibrator for High Signal-to-noise Sun-as-a-star Spectra}",
      journal = {\pasp},
         year = 2023,
        month = dec,
       volume = {135},
       number = {1054},
          eid = {125002},
        pages = {125002},
          doi = {10.1088/1538-3873/ad0b30},
archivePrefix = {arXiv},
       eprint = {2311.05129},
 primaryClass = {astro-ph.IM},
       adsurl = {https://ui.adsabs.harvard.edu/abs/2023PASP..135l5002R}
}

@ARTICLE{allesfitter-paper,
       author = {{G{\"u}nther}, Maximilian N. and {Daylan}, Tansu},
        title = "{Allesfitter: Flexible Star and Exoplanet Inference from Photometry and Radial Velocity}",
      journal = {\apjs},
         year = 2021,
        month = may,
       volume = {254},
       number = {1},
          eid = {13},
        pages = {13},
          doi = {10.3847/1538-4365/abe70e},
archivePrefix = {arXiv},
       eprint = {2003.14371},
 primaryClass = {astro-ph.EP},
       adsurl = {https://ui.adsabs.harvard.edu/abs/2021ApJS..254...13G}
}

@MISC{allesfitter-code,
 author = {{G{\"u}nther}, Maximilian~N. and {Daylan}, Tansu},
 title = "{Allesfitter: Flexible Star and Exoplanet Inference From Photometry and Radial Velocity}",
 howpublished = {Astrophysics Source Code Library},
 year = 2019,
 month = mar,
 archivePrefix = "ascl",
 eprint = {1903.003},
 adsurl = {http://adsabs.harvard.edu/abs/2019ascl.soft03003G}
}

@ARTICLE{Kipping2013,
       author = {{Kipping}, David M.},
        title = "{Efficient, uninformative sampling of limb darkening coefficients for two-parameter laws}",
      journal = {\mnras},
         year = 2013,
        month = nov,
       volume = {435},
       number = {3},
        pages = {2152-2160},
          doi = {10.1093/mnras/stt1435},
archivePrefix = {arXiv},
       eprint = {1308.0009},
 primaryClass = {astro-ph.SR},
       adsurl = {https://ui.adsabs.harvard.edu/abs/2013MNRAS.435.2152K}
}

@ARTICLE{Higson2019,
       author = {{Higson}, Edward and {Handley}, Will and {Hobson}, Mike and {Lasenby}, Anthony},
        title = "{Dynamic nested sampling: an improved algorithm for parameter estimation and evidence calculation}",
      journal = {Statistics and Computing},
         year = 2019,
        month = sep,
       volume = {29},
       number = {5},
        pages = {891-913},
          doi = {10.1007/s11222-018-9844-0},
archivePrefix = {arXiv},
       eprint = {1704.03459},
 primaryClass = {stat.CO},
       adsurl = {https://ui.adsabs.harvard.edu/abs/2019S&C....29..891H}
}

@ARTICLE{Wang2024,
       author = {{Wang}, Xian-Yu and {Rice}, Malena and {Wang}, Songhu and {Kanodia}, Shubham and {Dai}, Fei and {Logsdon}, Sarah E. and {Schweiker}, Heidi and {Teske}, Johanna K. and {Butler}, R. Paul and {Crane}, Jeffrey D. and {Shectman}, Stephen and {Quinn}, Samuel N. and {Kostov}, Veselin and {Osborn}, Hugh P. and {Goeke}, Robert F. and {Eastman}, Jason D. and {Shporer}, Avi and {Rapetti}, David and {Collins}, Karen A. and {Watkins}, Cristilyn N. and {Relles}, Howard M. and {Ricker}, George R. and {Seager}, Sara and {Winn}, Joshua N. and {Jenkins}, Jon M.},
        title = "{Single-star Warm-Jupiter Systems Tend to Be Aligned, Even around Hot Stellar Hosts: No T $_{eff}${\textendash}{\ensuremath{\lambda}} Dependency}",
      journal = {\apjl},
         year = 2024,
        month = sep,
       volume = {973},
       number = {1},
          eid = {L21},
        pages = {L21},
          doi = {10.3847/2041-8213/ad7469},
archivePrefix = {arXiv},
       eprint = {2408.10038},
 primaryClass = {astro-ph.EP},
       adsurl = {https://ui.adsabs.harvard.edu/abs/2024ApJ...973L..21W}
}

@ARTICLE{Hirano2011,
       author = {{Hirano}, Teruyuki and {Suto}, Yasushi and {Winn}, Joshua N. and {Taruya}, Atsushi and {Narita}, Norio and {Albrecht}, Simon and {Sato}, Bun'ei},
        title = "{Improved Modeling of the Rossiter-McLaughlin Effect for Transiting Exoplanets}",
      journal = {\apj},
         year = 2011,
        month = dec,
       volume = {742},
       number = {2},
          eid = {69},
        pages = {69},
          doi = {10.1088/0004-637X/742/2/69},
archivePrefix = {arXiv},
       eprint = {1108.4430},
 primaryClass = {astro-ph.EP},
       adsurl = {https://ui.adsabs.harvard.edu/abs/2011ApJ...742...69H}
}

@ARTICLE{Bruntt2010,
       author = {{Bruntt}, H. and {Bedding}, T.~R. and {Quirion}, P. -O. and {Lo Curto}, G. and {Carrier}, F. and {Smalley}, B. and {Dall}, T.~H. and {Arentoft}, T. and {Bazot}, M. and {Butler}, R.~P.},
        title = "{Accurate fundamental parameters for 23 bright solar-type stars}",
      journal = {\mnras},
         year = 2010,
        month = jul,
       volume = {405},
       number = {3},
        pages = {1907-1923},
          doi = {10.1111/j.1365-2966.2010.16575.x},
archivePrefix = {arXiv},
       eprint = {1002.4268},
 primaryClass = {astro-ph.SR},
       adsurl = {https://ui.adsabs.harvard.edu/abs/2010MNRAS.405.1907B}
}

@ARTICLE{Blanco2014iSpec,
       author = {{Blanco-Cuaresma}, S. and {Soubiran}, C. and {Heiter}, U. and {Jofr{\'e}}, P.},
        title = "{Determining stellar atmospheric parameters and chemical abundances of FGK stars with iSpec}",
      journal = {\aap},
         year = 2014,
        month = sep,
       volume = {569},
          eid = {A111},
        pages = {A111},
          doi = {10.1051/0004-6361/201423945},
archivePrefix = {arXiv},
       eprint = {1407.2608},
 primaryClass = {astro-ph.IM},
       adsurl = {https://ui.adsabs.harvard.edu/abs/2014A&A...569A.111B}
}

@ARTICLE{Blanco2014HighResolutionSpectralLibrary,
       author = {{Blanco-Cuaresma}, S. and {Soubiran}, C. and {Jofr{\'e}}, P. and {Heiter}, U.},
        title = "{The Gaia FGK benchmark stars. High resolution spectral library}",
      journal = {\aap},
         year = 2014,
        month = jun,
       volume = {566},
          eid = {A98},
        pages = {A98},
          doi = {10.1051/0004-6361/201323153},
archivePrefix = {arXiv},
       eprint = {1403.3090},
 primaryClass = {astro-ph.SR},
       adsurl = {https://ui.adsabs.harvard.edu/abs/2014A&A...566A..98B}
}

@ARTICLE{Blanco2019iSpec,
       author = {{Blanco-Cuaresma}, Sergi},
        title = "{Modern stellar spectroscopy caveats}",
      journal = {\mnras},
         year = 2019,
        month = jun,
       volume = {486},
       number = {2},
        pages = {2075-2101},
          doi = {10.1093/mnras/stz549},
archivePrefix = {arXiv},
       eprint = {1902.09558},
 primaryClass = {astro-ph.SR},
       adsurl = {https://ui.adsabs.harvard.edu/abs/2019MNRAS.486.2075B}
}

@ARTICLE{Doyle2014,
       author = {{Doyle}, Amanda P. and {Davies}, Guy R. and {Smalley}, Barry and {Chaplin}, William J. and {Elsworth}, Yvonne},
        title = "{Determining stellar macroturbulence using asteroseismic rotational velocities from Kepler}",
      journal = {\mnras},
         year = 2014,
        month = nov,
       volume = {444},
       number = {4},
        pages = {3592-3602},
          doi = {10.1093/mnras/stu1692},
archivePrefix = {arXiv},
       eprint = {1408.3988},
 primaryClass = {astro-ph.SR},
       adsurl = {https://ui.adsabs.harvard.edu/abs/2014MNRAS.444.3592D}
}

@ARTICLE{MasudaWinn2020,
       author = {{Masuda}, Kento and {Winn}, Joshua N.},
        title = "{On the Inference of a Star's Inclination Angle from its Rotation Velocity and Projected Rotation Velocity}",
      journal = {\aj},
         year = 2020,
        month = mar,
       volume = {159},
       number = {3},
          eid = {81},
        pages = {81},
          doi = {10.3847/1538-3881/ab65be},
archivePrefix = {arXiv},
       eprint = {2001.04973},
 primaryClass = {astro-ph.IM},
       adsurl = {https://ui.adsabs.harvard.edu/abs/2020AJ....159...81M}
}

@ARTICLE{OwenWu2017,
       author = {{Owen}, James E. and {Wu}, Yanqin},
        title = "{The Evaporation Valley in the Kepler Planets}",
      journal = {\apj},
         year = 2017,
        month = sep,
       volume = {847},
       number = {1},
          eid = {29},
        pages = {29},
          doi = {10.3847/1538-4357/aa890a},
archivePrefix = {arXiv},
       eprint = {1705.10810},
 primaryClass = {astro-ph.EP},
       adsurl = {https://ui.adsabs.harvard.edu/abs/2017ApJ...847...29O}
}

@ARTICLE{Ginzburg2018,
       author = {{Ginzburg}, Sivan and {Schlichting}, Hilke E. and {Sari}, Re'em},
        title = "{Core-powered mass-loss and the radius distribution of small exoplanets}",
      journal = {\mnras},
         year = 2018,
        month = may,
       volume = {476},
       number = {1},
        pages = {759-765},
          doi = {10.1093/mnras/sty290},
archivePrefix = {arXiv},
       eprint = {1708.01621},
 primaryClass = {astro-ph.EP},
       adsurl = {https://ui.adsabs.harvard.edu/abs/2018MNRAS.476..759G}
}

@ARTICLE{Gupta2019,
       author = {{Gupta}, Akash and {Schlichting}, Hilke E.},
        title = "{Sculpting the valley in the radius distribution of small exoplanets as a by-product of planet formation: the core-powered mass-loss mechanism}",
      journal = {\mnras},
         year = 2019,
        month = jul,
       volume = {487},
       number = {1},
        pages = {24-33},
          doi = {10.1093/mnras/stz1230},
archivePrefix = {arXiv},
       eprint = {1811.03202},
 primaryClass = {astro-ph.EP},
       adsurl = {https://ui.adsabs.harvard.edu/abs/2019MNRAS.487...24G}
}

@ARTICLE{Owen2019,
       author = {{Owen}, James E.},
        title = "{Atmospheric Escape and the Evolution of Close-In Exoplanets}",
      journal = {Annual Review of Earth and Planetary Sciences},
         year = 2019,
        month = may,
       volume = {47},
        pages = {67-90},
          doi = {10.1146/annurev-earth-053018-060246},
archivePrefix = {arXiv},
       eprint = {1807.07609},
 primaryClass = {astro-ph.EP},
       adsurl = {https://ui.adsabs.harvard.edu/abs/2019AREPS..47...67O}
}

@BOOK{Gray2005,
       author = {{Gray}, David F.},
        title = "{The Observation and Analysis of Stellar Photospheres}",
         year = 2005,
    publisher = {Cambridge University Press},
          doi = {10.1017/CBO9781316036570},
       adsurl = {https://ui.adsabs.harvard.edu/abs/2005oasp.book.....G}
}

@ARTICLE{Gagne2026,
       author = {{Gagn{\'e}}, Jonathan and {Moranta}, Leslie and {Faherty}, Jacqueline K. and {Curtis}, Jason Lee and {Bickle}, Thomas P. and {Couture}, Dominic and {Chiasson David}, Am{\'e}lie and {Christie}, Katie and {Lambier}, Samantha and {Leclerc}, Elise and {Poliquin}, Livia and {Belzile}, Danika and {Mamajek}, Eric E.},
        title = "{The Montreal Open Clusters and Associations (MOCA) Database: A Census of Nearby Associations, Open Clusters, and Young Substellar Objects within 500 pc of the Sun}",
      journal = {arXiv e-prints},
         year = 2026,
        month = feb,
          eid = {arXiv:2602.15695},
        pages = {arXiv:2602.15695},
          doi = {10.48550/arXiv.2602.15695},
archivePrefix = {arXiv},
       eprint = {2602.15695},
 primaryClass = {astro-ph.SR},
       adsurl = {https://ui.adsabs.harvard.edu/abs/2026arXiv260215695G}
}

@ARTICLE{Robbins2023,
       author = {{Robbins}, Grady and {Meisner}, Aaron M. and {Schneider}, Adam C. and {Burgasser}, Adam J. and {Kirkpatrick}, J. Davy and {Gagn{\'e}}, Jonathan and {Hsu}, Chih-Chun and {Moranta}, Leslie and {Casewell}, Sarah and {Marocco}, Federico and {Gerasimov}, Roman and {Faherty}, Jacqueline K. and {Kuchner}, Marc J. and {Caselden}, Dan and {Cushing}, Michael C. and {Alejandro}, Sherelyn and {Backyard Worlds: Cool Neighbors Collaboration}},
        title = "{CWISE J105512.11+544328.3: A Nearby Y Dwarf Spectroscopically Confirmed with Keck/NIRES}",
      journal = {\apj},
         year = 2023,
        month = nov,
       volume = {958},
       number = {1},
          eid = {94},
        pages = {94},
          doi = {10.3847/1538-4357/ad0043},
archivePrefix = {arXiv},
       eprint = {2310.09524},
 primaryClass = {astro-ph.SR},
       adsurl = {https://ui.adsabs.harvard.edu/abs/2023ApJ...958...94R}
}

@ARTICLE{Moranta2022,
       author = {{Moranta}, Leslie and {Gagn{\'e}}, Jonathan and {Couture}, Dominic and {Faherty}, Jacqueline K.},
        title = "{New Coronae and Stellar Associations Revealed by a Clustering Analysis of the Solar Neighborhood}",
      journal = {\apj},
         year = 2022,
        month = nov,
       volume = {939},
       number = {2},
          eid = {94},
        pages = {94},
          doi = {10.3847/1538-4357/ac8c25},
archivePrefix = {arXiv},
       eprint = {2206.04567},
 primaryClass = {astro-ph.SR},
       adsurl = {https://ui.adsabs.harvard.edu/abs/2022ApJ...939...94M}
}

@ARTICLE{Barragan2022_TOI560,
       author = {{Barrag{\'a}n}, O. and {Armstrong}, D.~J. and {Gandolfi}, D. and {Carleo}, I. and {Vidotto}, A.~A. and {Villarreal D'Angelo}, C. and {Oklop{\v{c}}i{\'c}}, A. and {Isaacson}, H. and {Oddo}, D. and {Collins}, K. and {Fridlund}, M. and {Sousa}, S.~G. and {Persson}, C.~M. and {Hellier}, C. and {Howell}, S. and {Howard}, A. and {Redfield}, S. and {Eisner}, N. and {Georgieva}, I.~Y. and {Dragomir}, D. and {Bayliss}, D. and {Nielsen}, L.~D. and {Klein}, B. and {Aigrain}, S. and {Zhang}, M. and {Teske}, J. and {Twicken}, J.~D. and {Jenkins}, J. and {Esposito}, M. and {Van Eylen}, V. and {Rodler}, F. and {Adibekyan}, V. and {Alarcon}, J. and {Anderson}, D.~R. and {Akana Murphy}, J.~M. and {Barrado}, D. and {Barros}, S.~C.~C. and {Benneke}, B. and {Bouchy}, F. and {Bryant}, E.~M. and {Butler}, R.~P. and {Burt}, J. and {Cabrera}, J. and {Casewell}, S. and {Chaturvedi}, P. and {Cloutier}, R. and {Cochran}, W.~D. and {Crane}, J. and {Crossfield}, I. and {Crouzet}, N. and {Collins}, K.~I. and {Dai}, F. and {Deeg}, H.~J. and {Deline}, A. and {Demangeon}, O.~D.~S. and {Dumusque}, X. and {Figueira}, P. and {Furlan}, E. and {Gnilka}, C. and {Goad}, M.~R. and {Goffo}, E. and {Guti{\'e}rrez-Canales}, F. and {Hadjigeorghiou}, A. and {Hartman}, Z. and {Hatzes}, A.~P. and {Harris}, M. and {Henderson}, B. and {Hirano}, T. and {Hojjatpanah}, S. and {Hoyer}, S. and {Kab{\'a}th}, P. and {Korth}, J. and {Lillo-Box}, J. and {Luque}, R. and {Marmier}, M. and {Mo{\v{c}}nik}, T. and {Muresan}, A. and {Murgas}, F. and {Nagel}, E. and {Osborne}, H.~L.~M. and {Osborn}, A. and {Osborn}, H.~P. and {Palle}, E. and {Raimbault}, M. and {Ricker}, G.~R. and {Rubenzahl}, R.~A. and {Stockdale}, C. and {Santos}, N.~C. and {Scott}, N. and {Schwarz}, R.~P. and {Shectman}, S. and {Raimbault}, M. and {Seager}, S. and {S{\'e}gransan}, D. and {Serrano}, L.~M. and {Skarka}, M. and {Smith}, A.~M.~S. and {{\v{S}}ubjak}, J. and {Tan}, T.~G. and {Udry}, S. and {Watson}, C. and {Wheatley}, P.~J. and {West}, R. and {Winn}, J.~N. and {Wang}, S.~X. and {Wolfgang}, A. and {Ziegler}, C.},
        title = "{The young HD 73583 (TOI-560) planetary system: two 10-M$_{{\ensuremath{\oplus}}}$ mini-Neptunes transiting a 500-Myr-old, bright, and active K dwarf}",
      journal = {\mnras},
         year = 2022,
        month = aug,
       volume = {514},
       number = {2},
        pages = {1606-1627},
          doi = {10.1093/mnras/stac638},
archivePrefix = {arXiv},
       eprint = {2110.13069},
 primaryClass = {astro-ph.EP},
       adsurl = {https://ui.adsabs.harvard.edu/abs/2022MNRAS.514.1606B}
}

@ARTICLE{Mamajek2008,
       author = {{Mamajek}, Eric E. and {Hillenbrand}, Lynne A.},
        title = "{Improved Age Estimation for Solar-Type Dwarfs Using Activity-Rotation Diagnostics}",
      journal = {\apj},
         year = 2008,
        month = nov,
       volume = {687},
       number = {2},
        pages = {1264-1293},
          doi = {10.1086/591785},
archivePrefix = {arXiv},
       eprint = {0807.1686},
 primaryClass = {astro-ph},
       adsurl = {https://ui.adsabs.harvard.edu/abs/2008ApJ...687.1264M}
}

@ARTICLE{ElMufti2023,
       author = {{El Mufti}, Mohammed and {Plavchan}, Peter P. and {Isaacson}, Howard and {Cale}, Bryson L. and {Feliz}, Dax L. and {Reefe}, Michael A. and {Hellier}, Coel and {Stassun}, Keivan and {Eastman}, Jason and {Polanski}, Alex and {Crossfield}, Ian J.~M. and {Gaidos}, Eric and {Kostov}, Veselin and {Wittrock}, Justin M. and {Villase{\~n}or}, Joel and {Schlieder}, Joshua E. and {Bouma}, Luke G. and {Collins}, Kevin I. and {Zohrabi}, Farzaneh and {Lee}, Rena A. and {Sohani}, Ahmad and {Berberian}, John and {Vermilion}, David and {Newman}, Patrick and {Geneser}, Claire and {Tanner}, Angelle and {Batalha}, Natalie M. and {Dressing}, Courtney and {Fulton}, Benjamin and {Howard}, Andrew W. and {Huber}, Daniel and {Kane}, Stephen R. and {Petigura}, Erik A. and {Robertson}, Paul and {Roy}, Arpita and {Weiss}, Lauren M. and {Behmard}, Aida and {Beard}, Corey and {Chontos}, Ashley and {Dai}, Fei and {Dalba}, Paul A. and {Fetherolf}, Tara and {Giacalone}, Steven and {Hill}, Michelle L. and {Hirsch}, Lea A. and {Holcomb}, Rae and {Lubin}, Jack and {Mayo}, Andrew and {Mo{\v{c}}nik}, Teo and {Akana Murphy}, Joseph M. and {Rosenthal}, Lee J. and {Rubenzahl}, Ryan A. and {Scarsdale}, Nicholas and {Stockdale}, Christopher and {Collins}, Karen and {Cloutier}, Ryan and {Relles}, Howard and {Tan}, Thiam-Guan and {Scott}, Nicholas J. and {Hartman}, Zach and {Matthews}, Elisabeth and {Ciardi}, David R. and {Gonzales}, Erica and {Matson}, Rachel A. and {Beichman}, Charles and {Bieryla}, Allyson and {Furlan}, E. and {Gnilka}, Crystal L. and {Howell}, Steve B. and {Ziegler}, Carl and {Brice{\~n}o}, C{\'e}sar and {Law}, Nicholas and {Mann}, Andrew W. and {Rabus}, Markus and {Johnson}, Marshall C. and {Christiansen}, Jessie and {Kreidberg}, Laura and {Berardo}, David Anthony and {Deming}, Drake and {Gorjian}, Varoujan and {Morales}, Farisa Y. and {Benneke}, Bj{\"o}rn and {Dragomir}, Diana and {Wittenmyer}, Robert A. and {Ballard}, Sarah and {Bowler}, Brendan P. and {Horner}, Jonathan and {Kielkopf}, John and {Liu}, Huigen and {Shporer}, Avi and {Tinney}, C.~G. and {Zhang}, Hui and {Wright}, Duncan J. and {Addison}, Brett C. and {Mengel}, Matthew W. and {Okumura}, Jack},
        title = "{TOI 560: Two Transiting Planets Orbiting a K Dwarf Validated with iSHELL, PFS, and HIRES RVs}",
      journal = {\aj},
         year = 2023,
        month = jan,
       volume = {165},
       number = {1},
          eid = {10},
        pages = {10},
          doi = {10.3847/1538-3881/ac9834},
archivePrefix = {arXiv},
       eprint = {2112.13448},
 primaryClass = {astro-ph.EP},
       adsurl = {https://ui.adsabs.harvard.edu/abs/2023AJ....165...10E}
}

@ARTICLE{Biddle2025Nature,
       author = {{Biddle}, Lauren I. and {Bowler}, Brendan P. and {Morgan}, Marvin and {Tran}, Quang H. and {Wu}, Ya-Lin},
        title = "{One-third of Sun-like stars are born with misaligned planet-forming disks}",
      journal = {\nat},
         year = 2025,
        month = aug,
       volume = {644},
       number = {8076},
        pages = {356-361},
          doi = {10.1038/s41586-025-09324-0},
archivePrefix = {arXiv},
       eprint = {2508.06488},
 primaryClass = {astro-ph.EP},
       adsurl = {https://ui.adsabs.harvard.edu/abs/2025Natur.644..356B}
}

@ARTICLE{Winter2025,
       author = {{Winter}, Andrew J. and {Benisty}, Myriam and {Izquierdo}, Andr{\'e}s F. and {Lodato}, Giuseppe and {Teague}, Richard and {Kimmig}, Carolin N. and {Andrews}, Sean M. and {Bae}, Jaehan and {Barraza-Alfaro}, Marcelo and {Cuello}, Nicol{\'a}s and {Curone}, Pietro and {Czekala}, Ian and {Facchini}, Stefano and {Fasano}, Daniele and {Hall}, Cassandra and {Hardiman}, Caitlyn and {Hilder}, Thomas and {Ilee}, John D. and {Fukagawa}, Misato and {Longarini}, Cristiano and {M{\'e}nard}, Fran{\c{c}}ois and {Orihara}, Ryuta and {Pinte}, Christophe and {Price}, Daniel J. and {Rosotti}, Giovanni and {Stadler}, Jochen and {Wilner}, David J. and {W{\"o}lfer}, Lisa and {Yen}, Hsi-Wei and {Yoshida}, Tomohiro C. and {Zawadzki}, Brianna},
        title = "{exoALMA. XVIII. Interpreting Large-scale Kinematic Structures as Moderate Warping}",
      journal = {\apjl},
         year = 2025,
        month = sep,
       volume = {990},
       number = {1},
          eid = {L10},
        pages = {L10},
          doi = {10.3847/2041-8213/adf113},
archivePrefix = {arXiv},
       eprint = {2507.11669},
 primaryClass = {astro-ph.EP},
       adsurl = {https://ui.adsabs.harvard.edu/abs/2025ApJ...990L..10W}
}

@ARTICLE{GJ3470b,
       author = {{Stef{\`a}nsson}, Gu{\dj}mundur and {Mahadevan}, Suvrath and {Petrovich}, Cristobal and {Winn}, Joshua N. and {Kanodia}, Shubham and {Millholland}, Sarah C. and {Maney}, Marissa and {Ca{\~n}as}, Caleb I. and {Wisniewski}, John and {Robertson}, Paul and {Ninan}, Joe P. and {Ford}, Eric B. and {Bender}, Chad F. and {Blake}, Cullen H. and {Cegla}, Heather and {Cochran}, William D. and {Diddams}, Scott A. and {Dong}, Jiayin and {Endl}, Michael and {Fredrick}, Connor and {Halverson}, Samuel and {Hearty}, Fred and {Hebb}, Leslie and {Hirano}, Teruyuki and {Lin}, Andrea S.~J. and {Logsdon}, Sarah E. and {Lubar}, Emily and {McElwain}, Michael W. and {Metcalf}, Andrew J. and {Monson}, Andrew and {Rajagopal}, Jayadev and {Ramsey}, Lawrence W. and {Roy}, Arpita and {Schwab}, Christian and {Schweiker}, Heidi and {Terrien}, Ryan C. and {Wright}, Jason T.},
        title = "{The Warm Neptune GJ 3470b Has a Polar Orbit}",
      journal = {\apjl},
         year = 2022,
        month = jun,
       volume = {931},
       number = {2},
          eid = {L15},
        pages = {L15},
          doi = {10.3847/2041-8213/ac6e3c},
archivePrefix = {arXiv},
       eprint = {2111.01295},
 primaryClass = {astro-ph.EP},
       adsurl = {https://ui.adsabs.harvard.edu/abs/2022ApJ...931L..15S}
}

@ARTICLE{GJ436b,
       author = {{Bourrier}, V. and {Zapatero Osorio}, M.~R. and {Allart}, R. and {Attia}, M. and {Cretignier}, M. and {Dumusque}, X. and {Lovis}, C. and {Adibekyan}, V. and {Borsa}, F. and {Figueira}, P. and {Gonz{\'a}lez Hern{\'a}ndez}, J.~I. and {Mehner}, A. and {Santos}, N.~C. and {Schmidt}, T. and {Seidel}, J.~V. and {Sozzetti}, A. and {Alibert}, Y. and {Casasayas-Barris}, N. and {Ehrenreich}, D. and {Lo Curto}, G. and {Martins}, C.~J.~A.~P. and {Di Marcantonio}, P. and {M{\'e}gevand}, D. and {Nunes}, N.~J. and {Palle}, E. and {Poretti}, E. and {Sousa}, S.~G.},
        title = "{The polar orbit of the warm Neptune GJ 436b seen with VLT/ESPRESSO}",
      journal = {\aap},
         year = 2022,
        month = jul,
       volume = {663},
          eid = {A160},
        pages = {A160},
          doi = {10.1051/0004-6361/202142559},
archivePrefix = {arXiv},
       eprint = {2203.06109},
 primaryClass = {astro-ph.EP},
       adsurl = {https://ui.adsabs.harvard.edu/abs/2022A&A...663A.160B}
}

@ARTICLE{HAT-P-11b,
       author = {{Sanchis-Ojeda}, Roberto and {Winn}, Joshua N.},
        title = "{Starspots, Spin-Orbit Misalignment, and Active Latitudes in the HAT-P-11 Exoplanetary System}",
      journal = {\apj},
         year = 2011,
        month = dec,
       volume = {743},
       number = {1},
          eid = {61},
        pages = {61},
          doi = {10.1088/0004-637X/743/1/61},
archivePrefix = {arXiv},
       eprint = {1107.2920},
 primaryClass = {astro-ph.EP},
       adsurl = {https://ui.adsabs.harvard.edu/abs/2011ApJ...743...61S}
}

@ARTICLE{WASP-107b,
       author = {{Rubenzahl}, Ryan A. and {Dai}, Fei and {Howard}, Andrew W. and {Chontos}, Ashley and {Giacalone}, Steven and {Lubin}, Jack and {Rosenthal}, Lee J. and {Isaacson}, Howard and {Batalha}, Natalie M. and {Crossfield}, Ian J.~M. and {Dressing}, Courtney and {Fulton}, Benjamin and {Huber}, Daniel and {Kane}, Stephen R. and {Petigura}, Erik A. and {Robertson}, Paul and {Roy}, Arpita and {Weiss}, Lauren M. and {Beard}, Corey and {Hill}, Michelle L. and {Mayo}, Andrew and {Mocnik}, Teo and {Murphy}, Joseph M. Akana and {Scarsdale}, Nicholas},
        title = "{The TESS-Keck Survey. IV. A Retrograde, Polar Orbit for the Ultra-low-density, Hot Super-Neptune WASP-107b}",
      journal = {\aj},
         year = 2021,
        month = mar,
       volume = {161},
       number = {3},
          eid = {119},
        pages = {119},
          doi = {10.3847/1538-3881/abd177},
archivePrefix = {arXiv},
       eprint = {2101.09371},
 primaryClass = {astro-ph.EP},
       adsurl = {https://ui.adsabs.harvard.edu/abs/2021AJ....161..119R}
}

@ARTICLE{Piaulet2021,
       author = {{Piaulet}, Caroline and {Benneke}, Bj{\"o}rn and {Rubenzahl}, Ryan A. and {Howard}, Andrew W. and {Lee}, Eve J. and {Thorngren}, Daniel and {Angus}, Ruth and {Peterson}, Merrin and {Schlieder}, Joshua E. and {Werner}, Michael and {Kreidberg}, Laura and {Jaouni}, Tareq and {Crossfield}, Ian J.~M. and {Ciardi}, David R. and {Petigura}, Erik A. and {Livingston}, John and {Dressing}, Courtney D. and {Fulton}, Benjamin J. and {Beichman}, Charles and {Christiansen}, Jessie L. and {Gorjian}, Varoujan and {Hardegree-Ullman}, Kevin K. and {Krick}, Jessica and {Sinukoff}, Evan},
        title = "{WASP-107b's Density Is Even Lower: A Case Study for the Physics of Planetary Gas Envelope Accretion and Orbital Migration}",
      journal = {\aj},
         year = 2021,
        month = feb,
       volume = {161},
       number = {2},
          eid = {70},
        pages = {70},
          doi = {10.3847/1538-3881/abcd3c},
archivePrefix = {arXiv},
       eprint = {2011.13444},
 primaryClass = {astro-ph.EP},
       adsurl = {https://ui.adsabs.harvard.edu/abs/2021AJ....161...70P}
}
\bibliographystyle{aasjournal}

\end{document}